\documentstyle[aps,prl,epsfig]{revtex}

\title{Relaxation and Noise in Chaotic Systems}

\author{Shmuel Fishman and Saar Rahav}
\address{Department of Physics, Technion, Haifa 32000, Israel}

\begin{document}

\draft
\date{Date: \today}
\maketitle
\begin{abstract}

For a class of idealized chaotic systems (hyperbolic systems) correlations decay exponentially in time. This result is
asymptotic and rigorous. The decay rate is related to the Ruelle-Pollicott resonances. Nearly all chaotic model systems, that
are studied by physicists, are not hyperbolic. For many such systems it is known that exponential decay takes place for
a long time. It may not be asymptotic, but it may persist for a very long time, longer than any time of experimental
relevance. In this review a heuristic method for calculation of this exponential decay of correlations in time is
presented. It can be applied to model systems, where there are no rigorous results concerning this exponential decay. It
was tested for several realistic systems (kicked rotor and kicked top) in addition to idealized systems (baker map and
perturbed cat map). The method consists of truncation of the evolution operator (Frobenius-Perron operator), and
performing all calculations with the resulting finite dimensional matrix. This finite dimensional approximation can be
considered as coarse graining, and is equivalent to the effect of noise. The  exponential decay rate of the chaotic system
is obtained when the dimensionality of the approximate evolution operator is taken to infinity, resulting in infinitely
fine resolution, that is equivalent to vanishing noise. The corresponding Ruelle-Pollicott resonances can be calculated
for many systems that are beyond the validity of the Ruelle-Pollicott theorem.

\end{abstract}

\section{Introduction}

The purpose of the lectures that are summarized in this review is to describe the behavior of ensembles of chaotic systems.
As in the case of statistical mechanics, the dynamics of ensembles turns out to be simpler then the one of individual
systems. For chaotic systems \cite{ott,arnold,haakeb,schuster,gaspard,dorfman,licht} the long time asymptotic behavior is rigorously 
known to exhibit exponential decay of correlations only
for a class of idealized systems (hyperbolic systems). Physical model systems do not belong to this class. In what follows
{\em nonrigorous} methods, that enable the exploration of the long time behavior of chaotic systems, are presented and their
application is demonstrated for several systems. The review is pedagogical and descriptive in nature, and is intended for an
overview of the subject. The reader should consult the references for the details and the precise statements.

The dynamics of the systems considered in this review is of Hamiltonian nature. The dynamics of the continuous systems 
is determined by the Hamilton equations:
\begin{eqnarray}
\label{hamilton}
\dot{p}=&-\frac{\partial {\cal H}}{\partial q}\\
\dot{q}=&\frac{\partial {\cal H}}{\partial p} \nonumber 
\end{eqnarray}
where ${\cal H}$ is the Hamiltonian, while  $q$ and $p$ are the position and momentum respectively. The phase space density 
$\rho({\bf x})$, where ${\bf x}=(q,p)$ are the phase space points, satisfies the Liouville theorem,
\begin{equation}
\label{li1}
\frac{d \rho}{dt}=0
\end{equation}
or 
\begin{equation}
\label{li2}
\frac{\partial \rho}{\partial t}=\{ {\cal H},\rho \} ,
\end{equation}
where $\{... \}$ are the Poisson brackets. The main issue that will be discussed is the way $\rho$ spreads in 
phase space for chaotic systems. 
For classical chaotic systems finer and finer structures develop for longer and longer times. These 
structures are reflected in $\rho$. Coarse graining, with some fixed scale of resolution, results in the truncation of the 
evolution of these very fine 
structures. On the finite (but arbitrarily small) resolution scale the equilibrium uniform phase space density is approached. In this 
review the asymptotic (in time) relaxation to this density will be discussed and in the end the coarse graining scale will be 
taken to $0$. The results differ from the ones obtained without coarse graining, since the limits of infinite time and vanishing 
coarse graining do not commute. The approach, where a finite resolution scale is used and then the limit where this scale tends to 
zero is taken,
is relevant for experimental realizations where the idealized classical description on the finest scales is destroyed resulting of 
the coupling to the environment. 

Maps are transformations of phase space in discrete time, denoted by $n$. A map ${\bf F}$ is a transformation 
\begin{equation}
\label{map}
{\bf x}_{n+1}={\bf F}({\bf x}_{n}). 
\end{equation}
We study maps since they are easier to handle analytically and numerically and they reproduce the most interesting results found for 
systems evolving continuously in time.

Maps can be derived from Hamiltonians of the form:
\begin{equation}
\label{map1}
{\cal H}=\frac{p^2}{2}+V(q)\sum_n \delta(t-n).
\end{equation}
For such maps the phase space area is preserved. Maps can also be defined with no reference to a Hamiltonian. The 
dynamics of phase space densities will be explored here for several area preserving maps: a. Kicked Rotor 
(Standard Map), b. Kicked Top, c. Arnold Cat Map and d. Baker Map.

\begin{figure}[b]
\begin{centering}
{\includegraphics[height=14cm,width=12cm]{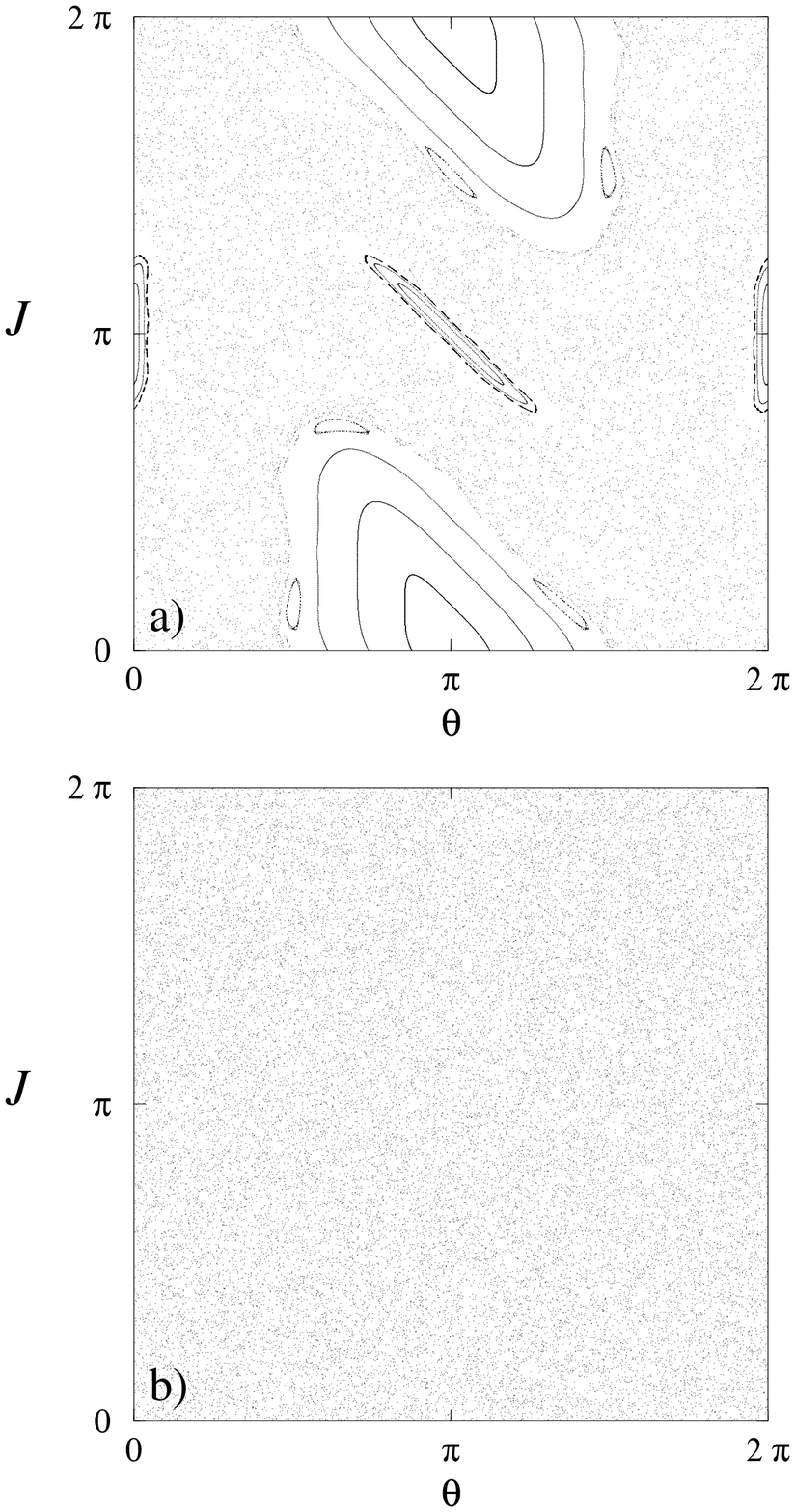}}
\caption{Phase space portraits of the weakly chaotic kicked rotor with stochasticity  parameter $K = 2$ (a), and of the strongly chaotic kicked rotor with $K = 10$ (b).\label{standard}}
\end{centering}
\end{figure}

{\bf a. Kicked Rotor (Standard Map)}\\ 
A planar rotor that is periodically kicked is modeled by the Hamiltonian \cite{ott,haakeb,licht} 
\begin{equation}
\label{kr1}
{\cal H}=\frac{J^2}{2}+K \cos\theta \sum_n \delta(t-n),
\end{equation}
where $\theta$ is the coordinate and $J$ is the conjugate momentum. The Hamilton equations (\ref{hamilton}) are:
\begin{eqnarray}
\label{kr2}
\dot{\theta}&=&J \\
\dot{J}&=&K \sin\theta \sum_n \delta(t-n). \nonumber
\end{eqnarray}
Integration with respect to time results in the Standard Map
\begin{eqnarray}
\label{kr3}
\theta_{n+1}&=&\theta_n+J_n \\
J_{n+1}&=&J_n+K \sin\theta_{n+1}, \nonumber
\end{eqnarray}
where $\theta_n$ and $J_n$ are the angle and angular momentum just after the $n$-th kick. The equations (\ref{kr3}) define a map 
of the form (\ref{map})
\begin{equation}
\label{kr4}
(\theta_{n+1},J_{n+1})={\bf F}(\theta_{n},J_{n}).
\end{equation}
It is easily checked that the map is area preserving. It becomes more chaotic as the stochasticity parameter $K$ increases. The phase space is plotted 
in Fig. \ref{standard}.  

A variant of this map provides a good description of driven laser cooled atoms \cite{raizen} and of beams deflected by 
dielectrics with a 
modulated index of refraction \cite{fischer}. 

\begin{figure}
\begin{centering}
{\includegraphics[height=12cm,width=8cm]{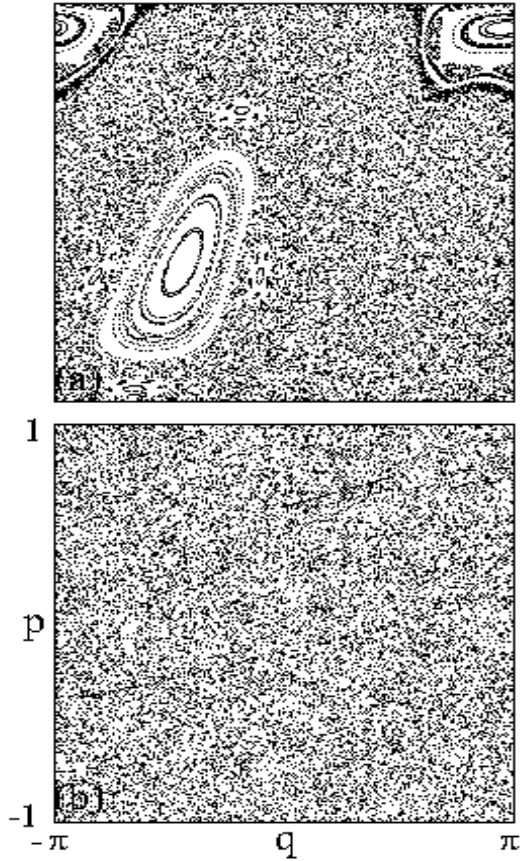}}
\caption{\label{haake1} Phase space portraits of the weakly chaotic kicked top with stochasticity parameter $\tau = 2.1$ (a), and of the strongly chaotic kicked top with $\tau = 10.2$ (b). (Fig. 1 of \protect\cite{haake}).}
\end{centering}
\end{figure}

{\bf b. Kicked Top}\\
A large spin can be described by a classical vector  
\begin{equation}
\label{kt1}
{\bf J}=j~(\sin\theta \cos \varphi, \sin\theta \sin \varphi, \cos\theta)
\end{equation}
where $\theta$  and $\varphi$ are the polar angles. The kicked top map is defined by the  transformation \cite{haakeb} 
\begin{equation}
\label{kt2}
{\bf F}=R_z(\tau \cos\theta)R_z(\beta_z)R_y(\beta_y),
\end{equation}
applied to ${\bf J}$ where $R_i(\beta)$ is the rotation around the axis $i$ by the angle $\beta$. The nonlinearity results 
of the dependence 
of a rotation on the angle $\theta$. The chaoticity of the map increases with $\tau$ that is the stochasticity parameter.
The canonical phase space variables for this map are $q=\varphi$ and $p=\cos\theta$.
The phase space is plotted in Fig. \ref{haake1}.  

{\bf c. Arnold Cat Map}\\
It is defined in the $[0,1] \times [0,1]$ square of the $(x,y)$ phase plane by  \cite{arnold,schuster}
\begin{eqnarray}
\label{cat1}
x_{n+1}=&x_n+y_n~~~~~  & \mbox{mod}~ 1 \\
y_{n+1}=&x_n+2y_n~~~~~ & \mbox{mod}~ 1. \nonumber
\end{eqnarray}
The evolution is demonstrated in \cite{arnold,schuster}.

\begin{figure}[tb]
\begin{centering}
{\includegraphics[height=8cm,width=8cm]{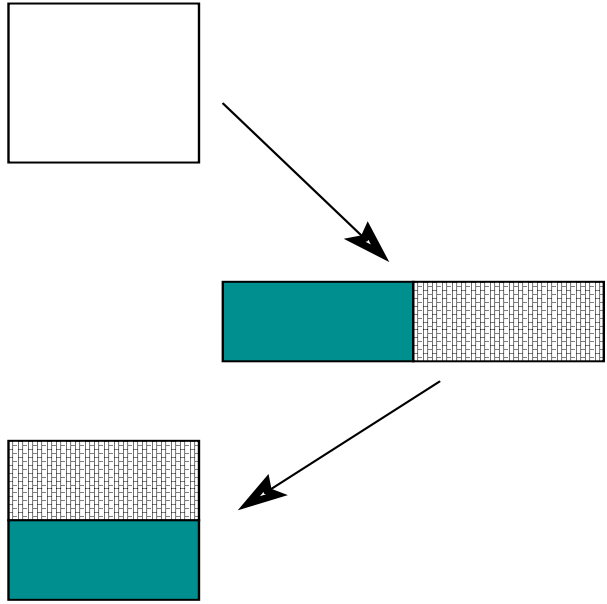}}
\caption{The baker map. \label{baker1}}
\end{centering}
\end{figure}

{\bf d. Baker map}\\
It is defined in the $[0,1] \times [0,1]$ square of the $(x,y)$ phase plane by \cite{schuster,dorfman,ott}
\begin{equation}
\label{baker}
(x_{n+1},y_{n+1})={\bf F}(x_n,y_n)=\left\{ \begin{array}{ll}
                 (2x_n,y_n/2)~~~~~~~~~~~~~~~~~~  &\mbox{ for $0 \leq x < \frac{1}{2}$}  \\
                 (2x_n-1,(y_n+1)/2)  &\mbox{ for $\frac{1}{2} \leq x < 1$}
                 \end{array}    
                 \right. 
\end{equation}
This transformation is demonstrated in Fig. \ref{baker1}. 

The systems (a) and (b) model physical problems. Their phase space is mixed, namely the dynamics in some parts is 
chaotic and in some parts it is regular. Systems (c) and (d) are very idealized. For these systems the motion is chaotic 
in the entire phase space. 

A system is chaotic if points, that are initially close, spread exponentially in phase space. For chaotic systems the separation 
between points grows with 
the number of steps as 
\begin{equation}
\label{lyap1}
\epsilon(n)=\epsilon(0)\Lambda^n,
\end{equation}
in the limit $n \rightarrow \infty$ and $\epsilon(0)\rightarrow 0$. The Lyapunov exponent is 
\begin{equation}
\label{lyap2}
\lambda=\ln \Lambda
\end{equation}
and $\Lambda$ is sometimes called the Lyapunov number. A system that evolves in a bounded region in phase space is 
called {\em chaotic} if its Lyapunov exponent $\lambda$  according to the definition (\ref{lyap1},\ref{lyap2}) is positive.
(More generally the Lyapunov numbers are the eigenvalues of the tangent map. The Lyapunov number defined by 
(\ref{lyap1}) is equal to the magnitude of the largest eigenvalue.) One can study the local expansion and contraction in the various 
directions. If for a map, for any point of phase space, in each direction, either expansion or contraction takes place, or 
in other words if none of the real parts of the local Lyapunov exponents vanishes, the system is called {\em hyperbolic}. 

Exponential spread of trajectories in phase space does {\em not} imply exponential decay of correlations in time. Such 
correlations may result of sticking to small structures in phase space, as is the case for the stadium billiard and for the 
Sinai billiard as well as for chaotic trajectories of mixed systems, such as the kicked rotor and the kicked top. For 
hyperbolic systems, such as the baker map, typically correlations in time decay exponentially.

Chaotic motion is reminiscent of a random walk. Therefore it is instructive to review the dynamics of the probability density of 
random walkers. Assume a random walk on a one dimensional lattice, with equal probability of $1/2$ to make a step to the 
left or to the right. The path of each walker is complicated but the evolution of a probability density of random walkers is 
simple. It is simple in particular in the continuum approximation (the limit of vanishing lattice spacing) where it is 
described by the diffusion equation
\begin{equation}
\label{dif1}
\frac{\partial }{\partial t}\rho(x,t)=D \frac{\partial^2}{\partial x^2} \rho(x,t),
\end{equation}
$D$ is the diffusion coefficient and $\rho(x,t)$ is the density of random walkers at the point $x$ at time $t$. 
Let us assume that the random walk is confined to an interval of length $s$, for example $-s/2 < x < s/2$. The current at the 
ends of the interval vanishes, resulting in the boundary conditions
\begin{equation}
\label{dif2}
\frac{\partial \rho}{\partial x}\left|_{x=\pm s/2}=0 \right.
\end{equation}
on (\ref{dif1}). This problem is similar to the standard quantum mechanical problem of a particle in an infinite square 
well, but with the boundary conditions (\ref{dif2}) and with no $i \hbar$ on the left hand side of (\ref{dif1}). Therefore the density of 
states can be expanded as:
\begin{equation}
\label{dif3}
\rho(x,t)=\sum_{k=0}^{\infty} a_k e^{-\gamma_k t}u_k(x),
\end{equation}
where
\begin{equation}
\label{dif4}
u_k=\left\{ \begin{array}{ll}
      \sqrt{\frac{2}{s}} \left[ \cos\frac{\pi x}{s}k+\left(\sqrt{\frac{1}{2}}-1 \right) \delta_{k0}\right]~~~~  &k~~ \mbox{even}  \\
      \sqrt{\frac{2}{s}} \sin\frac{\pi x}{s}k~~~~  &k~~ \mbox{odd}
                 \end{array}    
                 \right. 
\end{equation}
form an orthonormal basis and
\begin{equation}
\label{dif5}
\gamma_k=\left(\frac{\pi k}{s}\right)^2 D.
\end{equation}
The units of $a_k$ are of $1/\sqrt{\mbox{[length]}}$. 
The expansion coefficients $a_k$ are determined by the initial density $\rho(x,t=0)$.
The equilibrium density is 
\begin{equation}
\label{dif6}
\lim_{t \rightarrow \infty} \rho(x,t)=a_0 \sqrt{\frac{1}{s}}, 
\end{equation}
and the asymptotic approach to this density is
\begin{equation}
\label{dif7}
\rho(x,t)-\frac{a_0}{\sqrt{s}} \sim a_1 e^{-\gamma_1 t} u_1(x)
\end{equation}
with the relaxation rate $\gamma_1=\left(\frac{\pi }{s}\right)^2 D$. The density-density correlation function is
\begin{equation}
\label{dif8}
C(t)=\int_{-s/2}^{s/2} dx \rho(x,0) \rho(x,t)-a_0^2=\sum_{k=1}^{\infty}|a_k|^2 e^{-\gamma_k t} \sim |a_1|^2 e^{-\gamma_1 
t}.
\end{equation}
Hence also the correlations decay with the rate $\gamma_1$.

The example of the probability density of random walkers demonstrates that although the trajectories of specific random walkers are 
complicated their probability density $\rho$ follows simple dynamics. We turn now to study the phase space probability density $\rho({\bf x})$, 
that 
for Hamiltonian dynamics satisfies (\ref{li2}). The introduction of such probability densities, that are sufficiently smooth, is actually a 
coarse graining over some fine scale. It is an averaging over phase space of the same type that was performed for random walkers. 
For any given area preserving map ${\bf F}$ it is instructive to introduce the one step evolution operator $\hat{U}$, 
\begin{equation}
\label{U1}
\rho_{n+1}({\bf x})=\hat{U}\rho_n({\bf x}),
\end{equation}
where $\rho_n({\bf x})$ is the phase space density after $n$ steps of the map. If the operator is defined on a space of 
sufficiently 
smooth functions it is called the {\em Frobenius-Perron operator}. If ${\bf F}$ is area preserving and invertible, namely if 
${\bf 
F}^{-1}$ exists, then $\hat{U}$ is unitary. All eigenvalues of $\hat{U}$ are on the unit circle in the complex plane, they take 
the 
form $e^{-i2\pi \alpha}$, with real $\alpha$. The corresponding ``eigenfunctions'' are not square integrable. An example of such 
a 
function will be presented now \cite{berry}. Assume the map ${\bf F}$ has a periodic orbit of period $n$, that is:
\begin{eqnarray}
\label{sef1} 
{\bf x}_{j+1}&=&{\bf F}({\bf x}_j),~~~~~~~~~~~~~~~~~j=1,2,...n \\
{\bf x}_1&=&{\bf F}({\bf x}_n). \nonumber
\end{eqnarray}
The function 
\begin{equation}
\label{sef2}
\psi_{\alpha}({\bf x})=\sum_{j=1}^n \delta({\bf x}-{\bf x}_j)e^{i2\pi\alpha j}
\end{equation}
with $\alpha=l/n$, where $l$ is an integer, is an ``eigenfunction'' of $\hat{U}$  with the eigenvalue 
$e^{-i2\pi \alpha}$. Here $\psi_{\alpha}$ is called ``eigenfunction'' if 
\begin{equation}
\label{sef3}
\hat{U}\psi_{\alpha}=  e^{-i2\pi \alpha}\psi_{\alpha}
\end{equation}
and no requirement is made that it belongs to the space where $\hat{U}$ is defined. For example if $\hat{U}$ is defined on 
the space of square integrable functions, $\psi_{\alpha}$ of (\ref{sef2}) does not belong to this space and is not a function in the 
usual sense (it is a distribution). 

The operator $\hat{U}$ can be represented by an infinite dimensional matrix. In physics applications it is natural to use finite
dimensional approximations. Let us assume that a basis that is ordered by increased resolution is used (the basis states may be for
example sines and cosines or orthogonal polynomials). For the truncation of $\hat{U}$, namely a finite dimensional approximation,
the matrix is not unitary, and its eigenvalues are inside the unit circle in the complex plane. They may vary with the dimension of the
matrix. The natural question is: Do these eigenvalues approach the unit circle in the limit of an infinite dimensional matrix? It
will be demonstrated in what follows that this is {\em not} the case for chaotic systems and values {\em inside} the unit circle are approached
in this limit, in spite of the fact that the infinite dimensional matrix representing $\hat{U}$ is unitary. A heuristic 
justification for this
behavior was proposed  by F. Haake \cite{haake}. 
Chaotic systems, in contrast with regular ones, exhibit phase space structures on all scales. These structures are revealed 
during 
the evolution. The operator $\hat{U}$ couples the fine scales via its matrix elements that couple to states with high 
resolution. As a
result of the truncation, probability that was originally transferred to the fine scales is lost, resulting in nonunitarity of 
$\hat{U}$. Convergence of the eigenvalues to values inside the unit circle in the complex plane, in the limit of infinite 
dimension of the matrix, results of the asymptotic self similarity of the chaotic dynamics.
For regular motion, on the other hand, as the dimension of the matrix is increased the eigenvalues approach the unit circle 
\cite{haake} (see 
Sec. IV).

It will be shown that for chaotic systems the finite dimensional approximations of $\hat{U}$, in the limit of large dimension, 
describe the 
decay of correlations. The phase space density-density correlation function, in analogy with (\ref{dif8})  is  
\begin{eqnarray}
\label{cor1}
C(n)&=&\int d{\bf x} \rho({\bf x},0) \rho({\bf x},n)-\rho_{\infty}^2 \Omega \\ 
    &=&\int d{\bf x} \rho({\bf x},0) \hat{U}^n \rho({\bf x},0)-\rho_{\infty}^2 \Omega,   \nonumber
\end{eqnarray}
where $\Omega$ is the volume of the chaotic component in phase space.
The equilibrium density is $\rho_{\infty}=\lim_{n 
\rightarrow  \infty}\rho({\bf x},n)$ that is independent of position in phase space. 
A more general correlation function is:
\begin{equation}
\label{cor2}
C^{(A,B)}(n)=\int d{\bf x} A({\bf x}) \hat{U}^n B({\bf x}), 
\end{equation}
where we assumed 
\begin{equation}
\label{cor3}
\lim_{{n\rightarrow} \infty}\hat{U}^n A({\bf x})=\lim_{{n\rightarrow} \infty}\hat{U}^n B({\bf x})=0.  
\end{equation}
This can always be obtained by the subtraction of the asymptotic value. For simplicity it will be assumed that both $A$ and $B$ 
are real.

It is useful to study the Laplace transforms of the correlation functions. For this purpose we introduce the resolvent 
\begin{equation}
\label{cor4}
\hat{R}(z)=\frac{1}{z} \sum_{j=0}^{\infty}\hat{U}^j z^{-j}=\frac{1}{z-\hat{U}}
\end{equation}
where $z$ is a complex number. Since $\hat{U}$ is unitary the sum is convergent for $|z|>1$. This is analogous to the usual 
definition used in quantum mechanics:
\begin{equation}
\label{cor5}
\hat{R}(E)=\frac{1}{i \hbar}\int_0^{\infty}dt \hat{U} \exp \left( \frac{i}{\hbar}Et-\frac{\epsilon}{\hbar}t \right),
\end{equation}
where 
\begin{equation}
\label{cor6}
\hat{U}= \exp \left( -\frac{i}{\hbar}\hat{\cal H}t \right)
\end{equation}
is the evolution operator, leading to 
\begin{equation}
\label{cor7}
\hat{R}(E)=\frac{1}{E-\hat{\cal{H}}+i\epsilon}
\end{equation}
for $\epsilon>0$. The convergence of the integral (\ref{cor5}) requires  $\epsilon>0$. In analogy the convergence of the sum 
(\ref{cor4}) requires $|z|>1$ in (\ref{cor4}). The discrete Laplace (one sided Fourier) transform of the correlation function 
(\ref{cor2}) is 
\begin{equation}
\label{cor8}
\tilde{C}^{(A,B)}(z)=\sum_{n=0}^{\infty}C^{(A,B)}(n) z^{-n}=\int d{\bf x} A({\bf x})\left[z\hat{R}(z)\right] B({\bf x}), 
\end{equation}
as one finds from (\ref{cor4}). 

Let $\psi_i$ be an ``eigenfunction'' of $\hat{U}$ with the eigenvalue $z_i$,
\begin{equation}
\label{cor9}
\hat{U}\psi_i=  z_i\psi_i.
\end{equation}
By ``eigenfunction'' we mean here that it satisfies (\ref{cor9}), but it may not be a function in the usual sense (for example it 
may be a distribution). Then $z_i$ is a pole of the matrix elements of the resolvent as one can see from (\ref{cor4}). The 
correlation function, involving an ``eigenfunction'', takes the form
\begin{equation}
\label{cor10}
\tilde{C}^{(A,\psi_i)}(z)=\langle A|\psi_i\rangle \frac{z}{z-z_i},
\end{equation}
where the Dirac notation  
\begin{equation}
\label{cor11}
\langle A|B\rangle =\int d{\bf x} A^*({\bf x}) B({\bf x})
\end{equation}
is used ($A^*$ is the complex conjugate of $A$). Here $A$ and $B$ were assumed to be real. Let us expand $B({\bf x})$ in terms of 
the ``eigenfunctions'' of $\hat{U}$ as 
\begin{equation}
\label{cor12}
B({\bf x})=\sum_i b_i \psi_i({\bf x}).
\end{equation}
Then 
\begin{equation}
\label{cor13}
\tilde{C}^{(A,B)}(z)=\frac{r_i}{z-z_i}
\end{equation}
where the poles $z_i$ depend only on $\hat{U}$ while the residues 
\begin{equation}
\label{cor14}
r_i=b_i \langle A|\psi_i\rangle z,
\end{equation}
depend also on $A$ and $B$. 

Because of the unusual nature of the ``eigenfunctions'' $\psi_i$ the manipulations leading to (\ref{cor13}) are only heuristic. 
The Ruelle-Pollicott theorem justifies the expression (\ref{cor13}) for {\em hyperbolic} systems if $A$ and $B$ are typical smooth 
functions (for a precise statement of the theorem that is transparent for physicists see \cite{ruelle}). Moreover the theorem 
assures that 
$|z_i|<1$, except for the eigenvalue $z_0=1$ corresponding to the equilibrium density $\psi_0$, that is independent of ${\bf x}$. The 
$z_i$ are called the Ruelle-Pollicott resonances.

The existence of poles with $|z_i|<1$ implies the decay of correlations with the rate $\ln|z_i|$.   This results of 
\begin{equation}
\label{cor15}
C^{(A,B)}(n)=\frac{1}{2\pi i} \oint_{|z|=1} \frac{dz}{z} \tilde{C}^{(A,B)}(z)z^n
\end{equation}
and the application of the residue theorem to (\ref{cor13}). 

For any approximation of dimension $N$ of the Frobenius-Perron operator $\hat{U}$, that will be denoted by $\hat{U}^{(N)}$ in what follows, the eigenfunctions $\psi_i^{(N)}$, 
corresponding
to the eigenvalues $z_i^{(N)}$, are well defined. For hyperbolic systems, where the Ruelle-Pollicott theorem applies, it is
reasonable to assume that in the limit $N \rightarrow \infty$, the eigenvalues $z_i^{(N)}$ approach the Ruelle-Pollicott
resonances, for a typical choice of the basis.  Inspired by the Ruelle-Pollicott theorem the following heuristic scheme for the
calculation of the decay rates of correlations is proposed: \begin{enumerate}

\item Introduce an orthogonal basis where the basis states are ordered by resolution. These states, for example, may  be orthogonal 
polynomials
or trigonometric functions ($\cos 2\pi k x, \sin2\pi k x$) or exponentials $\exp{i2\pi k x}$.

\item Calculate the matrix elements of $\hat{U}$ in this basis.

\item Introduce a truncation of dimension $N$ of this matrix, that will be denoted as $\hat{U}^{(N)}$. 

\item Calculate the eigenvalues $z_i^{(N)}$ and the eigenfunctions $\psi_i^{(N)}$ of $\hat{U}^{(N)}$. The  eigenfunctions 
$\psi_i^{(N)}$ are finite linear combinations of the basis states, and therefore are smooth.  

\item Take the limit $N\rightarrow \infty$. In this limit $z_i^{(N)} \rightarrow z_i$ and $\psi_i^{(N)} \rightarrow \psi_i$.

\end{enumerate}

From the  Ruelle-Pollicott theorem it is expected, for hyperbolic systems, that the Ruelle-Pollicott resonances are obtained in this 
way, if a typical basis is used. The limiting functions $\psi_i$ are singular \cite{baker,baker1,baker2}. This scheme was applied
to various systems, including 
also to mixed systems \cite{haake,cat,top1,top2,rot,rot1}. Limiting eigenvalues $z_i$ satisfying $|z_i|<1$ were obtained in this way, 
implying exponential relaxation of 
correlations with the rate $\ln|z_i|$ that takes place for a very long time (for some nonhyperbolic systems it is known that eventually the 
correlations decay as a power-law, but the time required to obtain this power-law behavior may be too long for any physical 
relevance). 
Although the Hamiltonian dynamics of trajectories is reversible the phase space densities exhibit relaxation that is irreversible, and is similar to the 
behavior of probability densities of random walkers that follow
irreversible dynamics.

In Sec. II the heuristic scheme for the calculation of the  $z_i$ will be demonstrated for the baker map that is hyperbolic 
and consequently the Ruelle-Pollicott theorem applies. Exploration of a hyperbolic system that is a modification of the cat map is 
briefly mentioned in Sec. III. In Sec. IV the scheme will be applied to the kicked top and in Sec. V it 
will be applied to the kicked rotor, that are mixed systems, where the  Ruelle-Pollicott theorem is not valid. Also the limitations 
on
the validity of this scheme for such systems will be demonstrated. The main conclusions are summarized in Sec. VI.
The review will follow closely references \cite{haake,baker,rot}, that are marked by ** in the list of references. A detailed discussion on the 
relaxation of correlations in chaotic systems and on related topics can be found in \cite{gaspard,dorfman}. The scheme for the use of 
truncation in 
the calculation of the Ruelle-Pollicott resonances is discussed in a wider context in \cite{baker1}.

\section{The Frobenius-Perron Operator for the Baker Map, a Demonstration}

The heuristic prescription for the analysis of the Frobenius-Perron operator that was outlined in the Introduction will be 
demonstrated for the baker map (\ref{baker}), where all the results are exactly known. The analysis will make use of exact results obtained 
in 
\cite{baker}. 

The basis states that will be used are 
\begin{equation}
\label{leg1}
\langle xy|kl\rangle =\tilde{P}_k(x)\tilde{P}_l(y) 
\end{equation}
where
\begin{equation}
\label{leg2}
\tilde{P}_k(x)=\sqrt{2k+1}P_k(1-2x)=\frac{\sqrt{2k+1}}{k!}\frac{d^k}{dx^k}x^k(1-x)^k
\end{equation}
are the modified Legendre polynomials while $P_k$ are the Legendre polynomials. A basis of orthogonal polynomials is natural, since the order 
of the polynomial is not affected by the map (\ref{baker}).  The orthonormality of the basis
\begin{equation}
\label{leg3}
\langle kl|k'l'\rangle =\delta_{kk'}\delta_{ll'} 
\end{equation}
follows from the orthonormality of the modified Legendre polynomials
\begin{equation}
\label{leg4}
\int_0^1 dx \tilde{P}_k(x)\tilde{P}_{k'}(x)=\delta_{kk'}.
\end{equation}
The basis is 
naturally ordered 
by increased resolution, since this is a property of orthogonal polynomials. The action of $\hat{U}$ of (\ref{U1}) on any 
phase space density is
\begin{equation}
\label{ub1}
\hat{U}\rho({\bf x})=\rho({\bf F}^{-1}({\bf x})),
\end{equation}
since the points that are at ${\bf x}$ at time step $n+1$ were at ${\bf F}^{-1}({\bf x})$ the time step $n$. For the baker map 
(\ref{baker})
\begin{equation}
\label{ub2}
\rho({\bf F}^{-1}(x,y))
                  =\left\{ \begin{array}{ll}
                 (\rho(x/2,2 y)~~~~~~~~~~~  &\mbox{ for $0 \leq y < \frac{1}{2}$}  \\
                 (\rho((x+1)/2,2y-1)      &\mbox{ for $\frac{1}{2} \leq y < 1$}.
                 \end{array}
                 \right.
\end{equation}
From (\ref{ub1}) and (\ref{ub2}) one finds that the matrix elements of $\hat{U}$ are 
\begin{equation}
\label{lub3}
\langle kl|\hat{U}|k'l'\rangle=\frac{1}{2}\left[1+(-1)^{k+k'+l+l'}\right]I_{kk'}I_{l'l}
\end{equation}
where 
\begin{equation}
\label{ub4}
I_{kk'}=\int_0^1 dx \tilde{P}_k(x)\tilde{P}_{k'}(x/2).
\end{equation}
Using (\ref{leg2}) and integrating (\ref{ub4}) by parts $k$ times one finds that if $k>k'$,
\begin{equation}
\label{ub41}
I_{kk'}=0. 
\end{equation}
This results in the {\em nonrecurrence} property of $\hat{U}$ 
\begin{equation}
\label{ub5}
\langle kl|\hat{U}|k'l'\rangle =0~~~~~~~~~~~\mbox{for $k>k'$ or $l'>l$}.
\end{equation}
During the evolution, probability is transformed from states with $k'$ to states with $k$ only if $k \leq k'$, therefore after 
application of $\hat{U}$ the density becomes more uniform because the weight of lower order Legendre  polynomials is 
increased. This is 
expected since $x$ is the unstable direction, where stretching takes place, making the density more uniform. In the $y$ direction, 
on the other hand, the density is transformed from $l'$ to $l$ only if $l \geq l'$, therefore during the evolution, the weight of the high 
order Legendre polynomials increases. This results of the fact that $y$ is the stable direction, where contraction takes 
place, resulting in 
complexity that increases with time.

For $k \leq k'$ one finds \cite{baker}
\begin{equation}
\label{ub6}
I_{kk'}=\frac{[(2k+1)(2k'+1)]^{1/2}}{2^k}\sum_{l=0}^{k'-k}\left(-\frac{1}{2}\right)^l\frac{(k'+k+l)!}{(k'-k-l)!(2k+l+1)!l!}.
\end{equation}
In particular $I_{kk}=2^{-k}$ and the diagonal matrix elements are 
\begin{equation}
\label{ub7}
\langle kl|\hat{U}|kl\rangle =\frac{1}{2^{k+l}}.
\end{equation}

The matrix elements of  $\hat{U}^n$ can be calculated with the help of the nonrecurrence property (\ref{ub5}). Introducing 
the 
resolution of the identity one finds
\begin{equation}
\label{ub8}
\langle kl|\hat{U}^n|k'l'\rangle =\sum_{[k_i;l_i]} \langle kl|\hat{U}|k_1l_1\rangle \langle k_1l_1|\hat{U}|k_2l_2\rangle .....
~~~...\langle k_{n-1}l_{n-1}|\hat{U}|k'l'\rangle,
\end{equation}
where $[k_i;l_i] \equiv  [k_1,~k_2.....k_{n-1}; ~l_1,~l_2,....l_{n-1}]$ and the nonrecurrence property implies
$k \leq k_1 \leq k_2.....~~~~..\leq k'$ and $l \geq l_1  \geq l_2....~~~~...\geq l'$. In particular the diagonal matrix 
elements satisfy
\begin{equation}
\label{ub9}
\langle kl|\hat{U}^n|kl\rangle =\langle kl|\hat{U}|kl\rangle^n=\left(\frac{1}{2^{k+l}}\right)^n. 
\end{equation}
In the limit $n \rightarrow \infty$ the off-diagonal matrix elements are dominated by powers of the diagonal matrix elements, 
since the nonrecurrence property limits the number of non-diagonal matrix elements in (\ref{ub8}).

The diagonal matrix elements of the resolvent can be easily calculated with the help of (\ref{cor4}). For $|z|>1$ the sum in
(\ref{cor4}) is convergent and using (\ref{ub9}) one finds 
\begin{equation} 
\label{rb1}
\langle kl|\hat{R}(z)|kl\rangle =\sum_{n=0}^{\infty}z^{-(n+1)}\langle kl|\hat{U}|kl\rangle^n=\frac{1}{z-2^{-(k+l)}}.
\end{equation} 
These matrix elements are singular for $z=2^{-(k+l)}$, and except for $k=l=0$ all these singular points are inside the unit
circle in the complex plane. This is also the case for the off-diagonal matrix elements. Therefore the Ruelle-Pollicott
resonances are 
\begin{equation} 
\label{rb2} 
z_m=2^{-m} 
\end{equation} and their degeneracy is $m+1$. They are related to the
decay of correlations by (\ref{cor13}). They were obtained by analytic continuation from $|z|>1$ where the resolvent
is defined. One can show \cite{baker} that the matrix elements exhibit a cut at $|z|=1$ and the physically relevant poles are
on the Riemann sheet continued from $|z|>1$.

What happens if the Frobenius-Perron operator is restricted to $k,k' \leq k_{max}$ and $l,l' \leq l_{max}$ and $\hat{U}$ is
approximated by $\hat{U}^{(N)}$, an $N=k_{max}l_{max}$ dimensional matrix, resulting of the truncation of $\hat{U}$? The 
nonrecurrence
property holds for $\hat{U}^{(N)}$ and consequently (\ref{ub9}) holds. Therefore for the $N$ dimensional matrix $\hat{U}^{(N)}$ the
diagonal matrix elements, are zeros of the characteristic polynomial, that is of order $N$. To see this, note that by the 
Hamilton-Cayley theorem, $\hat{U}^{(N)}$ satisfies its characteristic polynomial, namely $\sum_{j=0}^N p_j \hat{U}^{(N)j}=0$, and calculate 
the diagonal matrix elements of this expression with the 
help of (\ref{ub9}). Consequently the diagonal matrix
elements are eigenvalues of $\hat{U}^{(N)}$, taking the values $2^{-m}$. The multiplicity of the eigenvalue $2^{-m}$ is $m+1$ if $m
\leq k_{max},l_{max}$. By a similarity transformation the matrix can be transformed to the canonical Jordan form. Another way to see
that the eigenvalues of $\hat{U}^{(N)}$ coincide with the diagonal matrix elements is by using the indices $l_{max}-l$ and
$l_{max}-l'$ instead of $l$ and $l'$. With these indices, $\hat{U}^{(N)}$ is upper triangular, therefore the eigenvalues coincide
with the diagonal matrix elements. The eigenvalues are independent of $k_{max}$ and $l_{max}$, therefore these are of the form
$2^{-m}$ also in the limit $k_{max} \rightarrow \infty$ and $l_{max}\rightarrow \infty$, in spite of the fact that $\hat{U}$ is
unitary. For $\hat{U}^{(N)}$ the eigenfunctions are finite combinations of Legendre polynomials and are therefore smooth. In the
limit $k_{max} \rightarrow \infty$ and $l_{max}\rightarrow \infty$ the right eigenfunctions are independent of $x$ and are
polynomials of $y$, while the left eigenfunctions tend to distributions \cite{baker}.

The main results that were obtained for the baker map are:

\begin{enumerate}

\item A basis that is ordered by increased resolution was introduced. A truncation of dimension $N$ in this basis was 
implemented. 

\item The Ruelle-Pollicott resonances were calculated from $\hat{U}^{(N)}$, the $N$ dimensional truncation of $\hat{U}$. As $N$ 
increases more eigenvalues are revealed. The values of the eigenvalues do not depend on the truncation dimension $N$, and therefore also 
in
the limit of infinite $N$ they remain at values that were obtained for finite $N$.

\item The poles of the matrix elements of the resolvent (\ref{cor4}) and (\ref{rb1}) were obtained by analytic continuation 
from $|z|>1$.

\item The eigenfunctions of  $\hat{U}$ are functions of $y$ only ($y$ is the stable direction).

\end{enumerate}

In this section the procedure for the calculation of  Ruelle-Pollicott resonances from $\hat{U}^{(N)}$, a truncated matrix 
approximating  $\hat{U}$, was 
demonstrated for a system where the results are known exactly. In the following sections this method will be applied to 
systems where there is no exact theory. 

The basis used here was of Legendre polynomials. If a basis of sines and cosines is used it is impossible to find the  
Ruelle-Pollicott resonances in this way, since each of the basis states collapses to $0$ after a finite number of 
applications 
of  $\hat{U}$ \cite{baker,com}. On the basis of the  Ruelle-Pollicott theorem, we believe that for a typical basis one does not 
encounter problems of this nature. 

\section{The Frobenius-Perron Operator of a Modified Cat Map}

For the Arnold cat map, defined by (\ref{cat1}), correlations in time decay faster than exponentially. A modification, where the 
function $f(x_{n+1})$, that is defined by 
\begin{equation}
\label{cat2} 
f(x)=\frac{K_0}{2\pi}\left[\cos(2\pi x)-\cos(4\pi x)\right], 
\end{equation}
is added to the equation for $y_{n+1}$ in (\ref{cat1}), is a  hyperbolic system if $K_0$ is sufficiently small \cite{ozorio}. The 
correlations for this system 
decay exponentially in time. It was studied with the help of the truncated Frobenius-Perron operator, according to the scheme that 
was 
outlined in the introduction. The leading Ruelle-Pollicott resonances were found with the help of a variational approach.

\section{The Frobenius-Perron Operator for the Kicked Top}

We turn now to apply the method where $\hat{U}$ is approximated by a finite dimensional matrix to mixed systems (where in some parts
of phase space the motion is chaotic and in other parts it is regular). The Ruelle-Pollicott theorem does not apply to such systems.
It will be demonstrated, however, that the Ruelle-Pollicott resonances are meaningful and describe the decay of
correlations in the chaotic component of mixed systems for a time that may be very long. Asymptotically power-law decay takes
place, due to sticking to regular islands in phase space.

In this section the kicked top~\cite{haakeb}, that was defined by (\ref{kt1}) and (\ref{kt2}), will be analyzed. It is a summary of results that 
were presented in \cite{haake} (see also \cite{top1} and \cite{top2}). The phase space is shown in Fig. \ref{haake1}.

\begin{figure}
\begin{centering}
{\includegraphics[height=9.5cm,width=10cm]{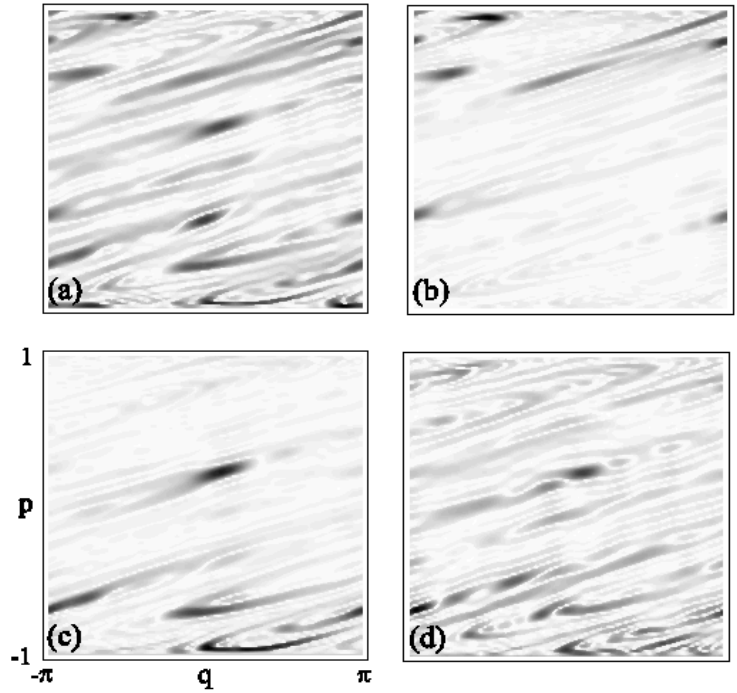}}
\caption{\label{haake4} The eigenfunctions corresponding to the eigenvalues $0.7696$ (a), $-0.3388\pm{\rm i}0.6243$ (b), $-0.0058\pm{\rm i}0.7080$
(c), $0.6480$ (d) of $\hat{U}^{(N)}$ for $\tau=10.2$, and $l_{{\rm max}} = 60$ (Fig. 4 of \protect\cite{haake}).}
\end{centering}
\end{figure}

\begin{figure}[tb]
\begin{centering}
{\includegraphics[height=9.5cm,width=10cm]{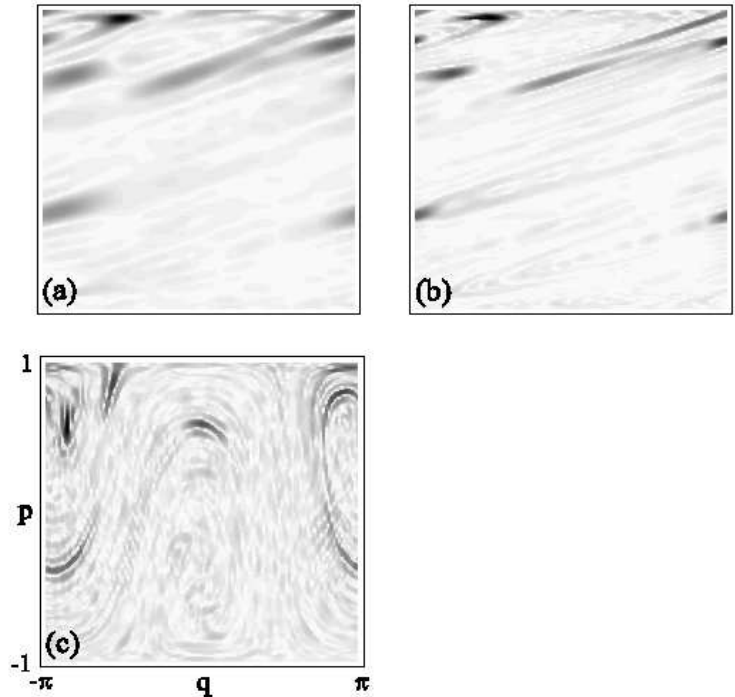}}
\caption{ \label{haake5} As phase-space resolution increases from $l_{{\rm max}}=30$ (a) to $l_{{\rm max}}=60$ (b), the eigenfunction of $\hat{U}^{(N)}$ for the eigenvalue  $-0.3388\pm {\rm i} 0.6243$ (for  $\tau=10.2$) gains  new structures on finer scales. The corresponding eigenfunction (c) of ${\hat{U}^{T(N)}}$ (resolution $l_{{\rm max}} = 60$) is localized at the same periodic orbit  as the eigenfunction (b) but with stable and unstable manifolds interchanged. See also figure \protect\ref{haake6}(a) for the periodic orbits, figure \protect\ref{haake6}(b) for the unstable and figure \protect\ref{haake6}(c) for the stable manifolds (Fig. 5 of  \protect\cite{haake}).}
\end{centering}
\end{figure}

\begin{figure}
\begin{centering}
{\includegraphics[height=12cm,width=12cm]{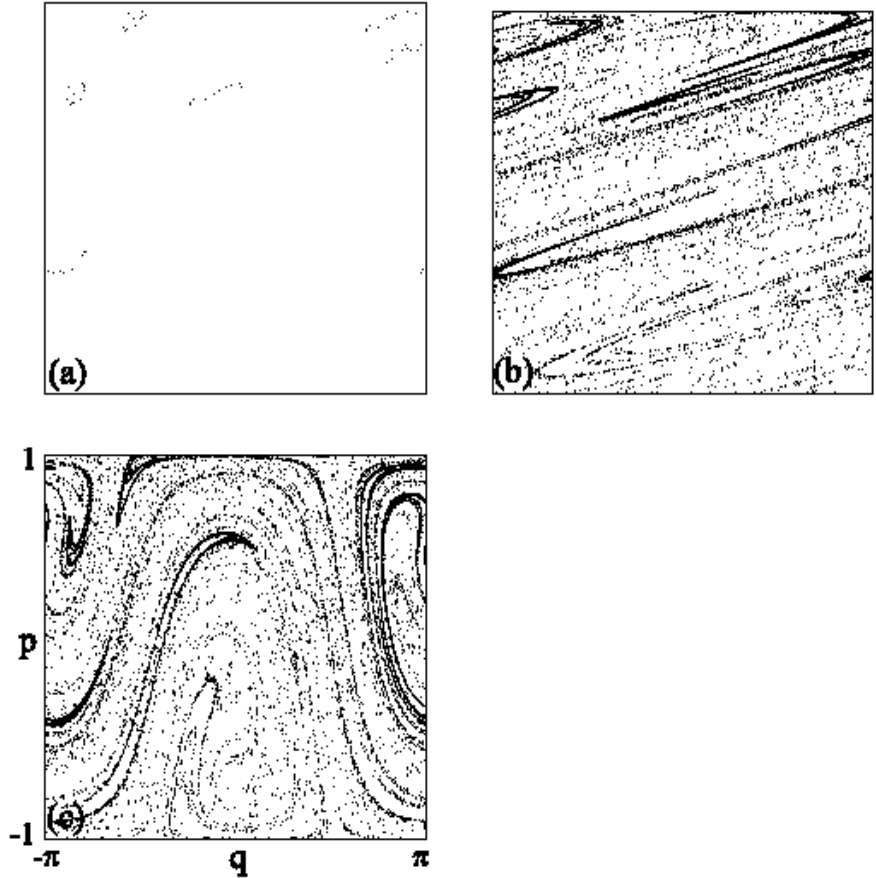}}
\caption{\label{haake6} (a): The $12$ orbits of primitive length $6$ that can be identified for the resonances $0.3550\pm {\rm i} 0.6199$ and
$-0.3388\pm {\rm i} 0.6243$ related to the eigenfunction shown in figures \protect\ref{haake4}(b) and \protect\ref{haake5}(b).  The
unstable manifolds of these orbits are shown in (b), and the stable manifolds in (c)  (Fig. 6 of \protect\cite{haake}).}
\end{centering}
\end{figure}

\begin{figure}
\begin{centering}
{\includegraphics[height=6cm,width=8cm]{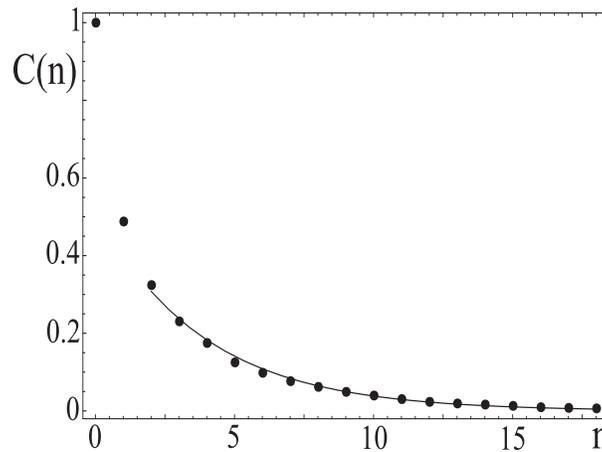}}
\caption{\label{haake7}The decay of $C(n)$ (dots) with $\rho(0)$ corresponding to the eigenfunction shown in figure \protect\ref{haake4}(b) (also \protect\ref{haake5}(b)). The numerical fit (line) yields a decay factor $0.7706$ (compared to $|z_i^{(N)}| \approx 0.7103$). (Fig. 7 of \protect\cite{haake}).}
\end{centering}
\end{figure}

A natural basis for this problem is of spherical harmonics
\begin{equation}
\label{sh}
Y_{lm}(\theta,\varphi)=(-1)^m \sqrt{\frac{2l+1}{4\pi} \frac{(l-m)!}{(l+m)!}}P_l^m(\cos\theta)e^{im\varphi}.
\end{equation}
Alternatively one can choose a basis of real functions with $e^{im\varphi}$ replaced by $\sin m\varphi$ and $\cos m\varphi$, resulting 
in a real matrix for $\hat{U}$. It is naturally ordered by increased resolution. A truncation at $l=l_{max}$ is introduced. The 
dimension of the truncated matrix  $\hat{U}^{(N)}$, approximating  $\hat{U}$ is $N=(l_{max}+1)^2$. As $l_{max}$ is increased the 
eigenvalues of $\hat{U}^{(N)}$ converge to values inside the unit circle in the complex plane, as can be seen from Table \ref{tableA}. 
\begin{table}
\[
\begin{array}{rrrr} \hline\hline
l_{{\rm max}} = 30 & l_{{\rm max}} = 40 & l_{{\rm max}} = 50 & l_{{\rm max}} = 60 \\ \hline
0.7700 & 0.7688 & 0.7523 & 0.7696 \\ \hline
\begin{array}{r} 0.3075\\ \pm {\rm i}\; 0.5740\end{array} &
\begin{array}{r} 0.3429\\ \pm {\rm i}\; 0.6140\end{array} &
\begin{array}{r} 0.3523\\ \pm {\rm i}\; 0.6211\end{array} &
\begin{array}{r} 0.3550\\ \pm {\rm i}\; 0.6199\end{array} \\ \hline
\begin{array}{r} -0.3170\\ \pm {\rm i}\; 0.6003\end{array} &
\begin{array}{r} -0.3348\\ \pm {\rm i}\; 0.6272\end{array} &
\begin{array}{r} -0.3444\\ \pm {\rm i}\; 0.6283\end{array} &
\begin{array}{r} -0.3388\\ \pm {\rm i}\; 0.6243\end{array} \\ \hline
\begin{array}{r} -0.0042\\ \pm {\rm i}\; 0.7161\end{array} &
\begin{array}{r} -0.0002\\ \pm {\rm i}\; 0.7133\end{array} &
\begin{array}{r} -0.0100\\ \pm {\rm i}\; 0.6930\end{array} &
\begin{array}{r} -0.0058\\ \pm {\rm i}\; 0.7080\end{array} \\ \hline
-0.7025 & -0.7228 & -0.7155 & -0.7165 \\ \hline
0.6544 & 0.6230 & 0.6495 & 0.6480 \\ \hline
\dots & -0.5619 & -0.5753 & -0.5667 \\ \hline\hline
\end{array}
\]
\caption{Eigenvalues of $\hat{U}^{(N)}$ for $\tau=10.2$ truncated at $l_{{\rm max}} =30, 40, 50$ and $60$ (Table I of \protect\cite{haake}).}
\label{tableA}
\end{table}
The 
corresponding eigenfunctions are presented in Fig. \ref{haake4}. 
Dark-shaded regions in phase space indicate large 
amplitudes of the eigenfunctions. In Figs. \ref{haake5} (a) and (b) 
it is 
demonstrated that as $l_{max}$ increases finer details of the eigenfunctions are revealed. This is a result of increased resolution. 
Comparison with Fig. \ref{haake6} (b)
demonstrates that the eigenfunctions are large and tend to be uniform on the 
unstable manifold. In Fig. \ref{haake7} 
the density-density correlation function (proportional to (\ref{cor1})) is 
plotted. 
It is demonstrated that also here the Ruelle-Pollicott resonances describe the decay of correlations for a very long time.

\begin{figure}
\begin{centering}
{\includegraphics[height=10cm,width=8cm]{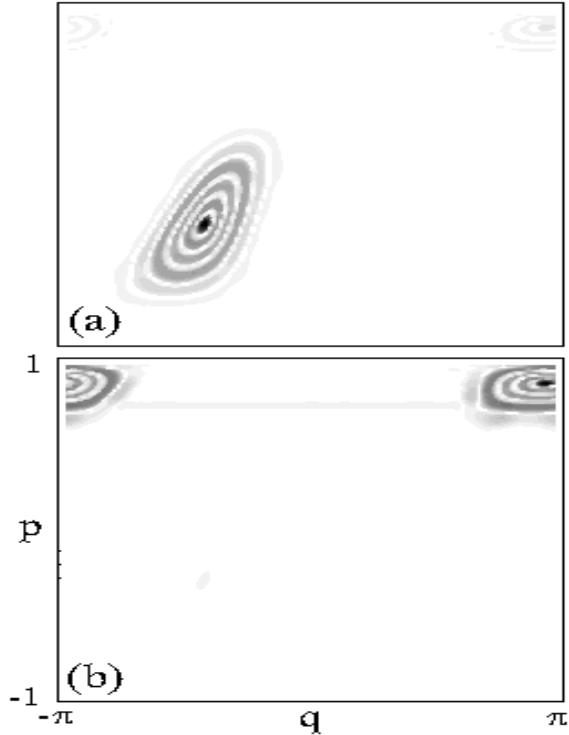}}
\caption{\label{haake3} Eigenfunctions of $\hat{U}^{(N)}$ with $l_{{\rm max}} = 60$ that are localized on elliptic islands shown in Fig. 2 (a) for the weakly chaotic case $\tau=2.1$. The corresponding eigenvalues $0.999976$ (a) and $0.999974$ (b) are almost at unity (Fig. 3 of \protect\cite{haake}).}
\end{centering}
\end{figure}

The behavior of the eigenfunctions localized in regular regions and the corresponding eigenvalues is very different. Eigenfunctions 
that are localized in islands of regular motion presented in Fig. \ref{haake1} (a) are depicted in Fig. \ref{haake3}.
The 
eigenfunctions follow elliptic orbits inside the regular island and therefore do not involve the resolution of very fine structures. 
The corresponding eigenvalues approach the unit circle in the complex plane as $l_{max}$ increases. 
 
It is instructive to examine the evolution operator  $\hat{U}^{-1}$ corresponding to the inverse map ${\bf F}^{-1}$. 
Unitarity implies 
\begin{equation}
\label{top1}
\hat{U}^{-1}=\hat{U}^{\dagger}.
\end{equation}
A basis can be chosen so that $\hat{U}$ is real and $\hat{U}^{\dagger}=\hat{U}^T$, where $T$ denotes transpose. The eigenvalues of 
$\hat{U}^T$ are equal to the ones of $\hat{U}$. Introducing the truncation one finds 
\begin{equation}
\label{top2}
\hat{U}^{T(N)}=\hat{U}^{(N)T},
\end{equation}
where $\hat{U}^{T(N)}$ is the truncated $\hat{U}^T$ and $\hat{U}^{(N)T}$ is the transpose of $\hat{U}^{(N)}$, the truncated 
$\hat{U}$. The eigenfunctions of $\hat{U}^{T(N)}$ are localized on the unstable manifold of ${\bf F}^{-1}$, that is the stable 
manifold 
of ${\bf F}$,
as is clearly demonstrated comparing Fig. \ref{haake5} (c) with Fig. \ref{haake6} (c). By eigenfunctions we meant so far the right eigenfunctions. The 
left eigenfunctions  of 
$\hat{U}$ (and $\hat{U}^{(N)}$)   
are the right eigenfunctions of $\hat{U}^{T}$ (and  $\hat{U}^{T(N)}$).

The main results that were obtained for the kicked top are:

\begin{enumerate}

\item As $N$ or $l_{max}$ increases the eigenvalues of $\hat{U}^{(N)}$, corresponding to eigenfunctions  localized in the chaotic 
component of phase space, converge to values inside the unit circle. Unlike the case of the baker map, the eigenvalues $z_i^{(N)}$
depend on $N$ and $\lim_{N \rightarrow \infty} z_i^{(N)}=z_i$ with $|z_i|<1$, except the unit eigenvalue corresponding to the 
equilibrium 
density. 

\item These eigenvalues determine the decay of correlations in the chaotic component of phase space. 

\item In the chaotic component of phase space the right eigenfunctions of the truncated Frobenius-Perron operator $\hat{U}^{(N)}$
are localized on the unstable manifold of ${\bf F}$, and tend to be uniform along this manifold. The left eigenfunctions are 
localized on the stable manifold. These manifolds are not as simple as for the baker map.

\item The eigenvalues of $\hat{U}^{(N)}$ corresponding to eigenfunctions localized in regular regions approach the unit circle as  
$N$ or $l_{max}$ increases. 

\end{enumerate}

\section{The Frobenius-Perron Operator for the Kicked Rotor}

In this section the decay of correlations in time will be studied for the kicked rotor, defined by the Hamiltonian 
(\ref{kr1}), and 
its dynamics is given by the standard map (\ref{kr3}) \cite{ott,haakeb,licht}. The phase space is presented in Fig. \ref{standard}. It is a mixed 
system (where in some regions of phase space the motion is chaotic and in other regions it is regular). In this section we summarize 
results that were presented in \cite{rot} (see also \cite{rot1}, \cite{balescu} and \cite{cat}).

For sufficiently large values of the stochasticity parameter $K$ (that takes a typical value), the spread of angular momentum is diffusive, to a good approximation, 
namely
\begin{equation}
\label{dkr1}
\left<(J_n-J_0)^2\right> \approx 2 D n,
\end{equation}
for large $n$. The average is over the initial points  and is denoted by $\left<.....\right>$. A very crude way to get this 
result 
is to iterate (\ref{kr3}) $n$ times to obtain 
\begin{equation}
\label{dkr2}
J_n-J_0=K\sum_{i=1}^n \sin\theta_i,
\end{equation}
then squaring and averaging with the assumption of absence of angular correlations 
\begin{equation}
\label{dkr3}
\left<\sin\theta_i \sin\theta_j\right>=\frac{1}{2}\delta_{ij},
\end{equation}
one obtains (\ref{dkr1}) with $D=\frac{1}{4}K^2$. A more careful calculation, that takes some of the angular correlations into 
account results in \cite{RW}
\begin{equation}
\label{dkr4}
D(K)=\frac{K^{2}}{4} \left(1-2J_{2}(K)+... \rule{0mm}{4mm} \right).
\end{equation}
It is actually an expansion in powers of $1/\sqrt{K}$. Because of the mixed nature of phase space it is expected that 
diffusion may not take place asymptotically in time \cite{Zaslav}. Moreover, there are values of $K$ (near integer multiples of 
$2\pi$) where 
acceleration in momentum is found for some initial conditions, resulting in $\left<(J_n-J_0)^2\right> \sim n^2$. In what follows 
the Frobenius-Perron operator will be used to study whether correlations in angular momentum decay as expected for a true 
diffusion 
process, described in the Introduction (see (\ref{dif1})-(\ref{dif8})). We will analyze also the decay of angular 
correlations. 

The Frobenius-Perron operator, corresponding to the map (\ref{kr3}) on the torus
\begin{equation}
\ \ \ (0 \leq J < 2\pi s)\\
\label{fpr1}
\end{equation}
\[
(0 \leq \theta < 2\pi),
\]
where $s$ is integer, will be studied. For the usual kicked rotor $s \rightarrow \infty $. Noise is added to the standard map by 
the addition of a random variable $\xi_n$ with variance $\sigma^2$ on the right hand side of the equation for $\theta_{n+1}$ in 
(\ref{kr3}) following \cite{RW}. The noise leads to truncation of the Frobenius-Perron operator. 
The natural basis for the analysis is 
\begin{equation}
\label{fpr2}
\langle J\theta|km \rangle =\frac{1}{\sqrt{2\pi}}\frac{1}{\sqrt{2\pi s}} \exp (im\theta)\exp \left( i\frac{kJ}{s} \right).
\end{equation}
where $k$ and $m$ are integers. Note that the functions $\langle J\theta|k,m=0 \rangle$ form the basis of eigenstates of the
diffusion operator in the angular momentum $J$. In this basis, in presence of noise, the Frobenius-Perron operator is
\begin{equation}
\label{fpr3}
\langle km|\hat{U}^{(\sigma)}|k'm'\rangle =J_{m-m'}\left(\frac{k'K}{s}\right)\exp 
\left(-\frac{\sigma^{2}}{2}m^{2}\right)\delta_{k-k',m 
s}.
\end{equation}
For $\sigma \neq 0$ the operator is not unitary. The effective truncation is $m<1/{\sigma},~|k-k'|<s/{\sigma},~|m-m'|<k'K/s$. 
The 
$\sigma \rightarrow 0$ limit , corresponding to the $N \rightarrow \infty $ limit of the previous sections, will be taken in 
the end of 
calculation. The analysis will be performed for 
\begin{eqnarray}
\label{fpr4}
s &\gg& 1 \\
K &\gg& 1  \nonumber \\ 
\frac{kK}{s} &\ll& 1 \nonumber
\end{eqnarray}
and the limit $\sigma \rightarrow 0$ will be taken in the end of the calculation. 

The Ruelle resonances are identified from the matrix elements of the resolvent (\ref{cor4}). It is useful to define a variant 
of the resolvent
\begin{equation}
\label{fpr5}
\hat{R}'(z)=\sum_{j=0}^{\infty}\hat{U}^j z^{j}=\frac{1}{1-z\hat{U}}
\end{equation}
that is related to $\hat{R}(z)$ via
\begin{equation}
\label{fpr6}
\frac{1}{z}\hat{R}'\left(\frac{1}{z}\right)=\hat{R}(z).
\end{equation}
Also the matrix elements of $\hat{R}$ and $\hat{R}'$ satisfy (\ref{fpr6}).
Let 
\begin{equation}
\label{fpr7}
R_{12}=\langle k_{1}m_{1}|\hat{R}(z)|k_{2}m_{2} \rangle
\end{equation}
and 
\begin{equation}
\label{fpr8}
R^{'}_{12}=\langle k_{1}m_{1}|\hat{R}'(z)|k_{2}m_{2} \rangle
\end{equation}
be the matrix elements of $\hat{R}$ and $\hat{R}'$ respectively. Therefore a singularity of $R_{12}$ at $z_c$ implies a 
singularity of $R^{'}_{12}$ at $1/z_c$ and vice versa. 
The series (\ref{cor4}) for $\hat{R}(z)$ converge for $|z|>1$ while the series (\ref{fpr5}) for $\hat{R}'(z)$ converge 
for  $|z|<1$. Therefore the Ruelle-Pollicott resonance (located inside) that is closest to the unit circle corresponds to the 
singularity 
of $R^{'}_{12}$ that is closest to the unit circle (located outside). This singularity is the radius of convergence 
of the series
\begin{equation}
\label{fpr9}
R^{'}_{12}=\sum_{j=0}^{\infty}a_{j}z^{j},
\end{equation}
where
\begin{equation}
\label{fpr10}
a_{j}=\langle k_{1}m_{1}|\hat{U}^{j}|k_{2}m_{2} \rangle.
\end{equation}
According to the Cauchy-Hadamard theorem (see ~\cite{CA})
the inverse of the radius of convergence is given by
\begin{equation}
\label{fpr11}
r^{-1}=\lim_{j \rightarrow \infty}\sup \sqrt[j]{|a_{j}|}.
\end{equation}
and asymptotically
\begin{equation}
\label{fpr12}
{|a_{j}|} \sim \frac{\mbox{const.}}{r^j}.
\end{equation}
Since the radius of convergence is the singularity of $R'_{12}(z)$ that is the closest to the unit circle it satisfies $r=1/z_c$
and $\sqrt[j]{a_{j}} \rightarrow z_c$ in the limit $j \rightarrow \infty $ (at least for some subsequence of $\{j\}$). 

The limit of the series (\ref{fpr10}) was found in the leading order in perturbation theory in the small parameters implied by 
(\ref{fpr4}). 
The calculation can be performed separately for the various values of $k$, resulting 
in
\begin{equation}
\label{krc1n}
z_{k}=\exp\left(-\frac{k^{2}K^{2}}{4s^{2}}\left(1-2J_{2}(K)e^{-\sigma^{2}}\rule{0mm}{4mm} \right) \right),
\end{equation}
corresponding to diffusion modes in presence of noise.
Taking the limit $\sigma \rightarrow 0 $ one finds
\begin{equation}
\label{krc11}
z_k=e^{-k^2 D(K)/s^2}
\end{equation}
where $D(K)$ is given by (\ref{dkr4}). This relaxation is similar to the one that is found for usual diffusion (see 
(\ref{dif1}-\ref{dif8})).
The relaxation rates are
\begin{equation}
\label{krc1}
\gamma_k=\frac{k^2}{s^2}D(K),
\end{equation}
that correspond to (\ref{dif5}).
These describe the decay of correlations in angular momentum $J$. The equilibrium density corresponds to $z_0$. Exploring the subspace 
involving $|k=0,m \rangle$, one finds that the Ruelle-Pollicott resonances, corresponding to the slowest mode of the relaxation of correlations 
in angle, take four values
\begin{equation}
\label{krc3}
\pm\tilde{z},~~~~~~~~~~~~\pm i\tilde{z}
\end{equation}
with
\begin{equation}
\label{krc4n}
\tilde{z}=\sqrt{\left|J_{2m^*}(m^* K)\right|\exp{\left(-\sigma^2 m^{*2}/2\right)}}.
\end{equation}
Taking the limit $\sigma \rightarrow 0 $ one finds 
\begin{equation}
\label{krc4}
\tilde{z}=\sqrt{\left|J_{2m^*}(m^* K)\right|},
\end{equation}
where $m^*$ is the integer $m$ that maximizes $\left|J_{2m}(m K)\right|$ for a given value of $K$. A similar result was found in 
\cite{cat} by a variational approach.
The corresponding relaxation rate is
\begin{equation}
\label{krc5}
\tilde{\gamma}=\ln \tilde{z}.
\end{equation}

\begin{figure}[tb]
\begin{center}
\begin{minipage}{7.01cm}
\centerline{\epsfxsize=7.0cm \epsfbox{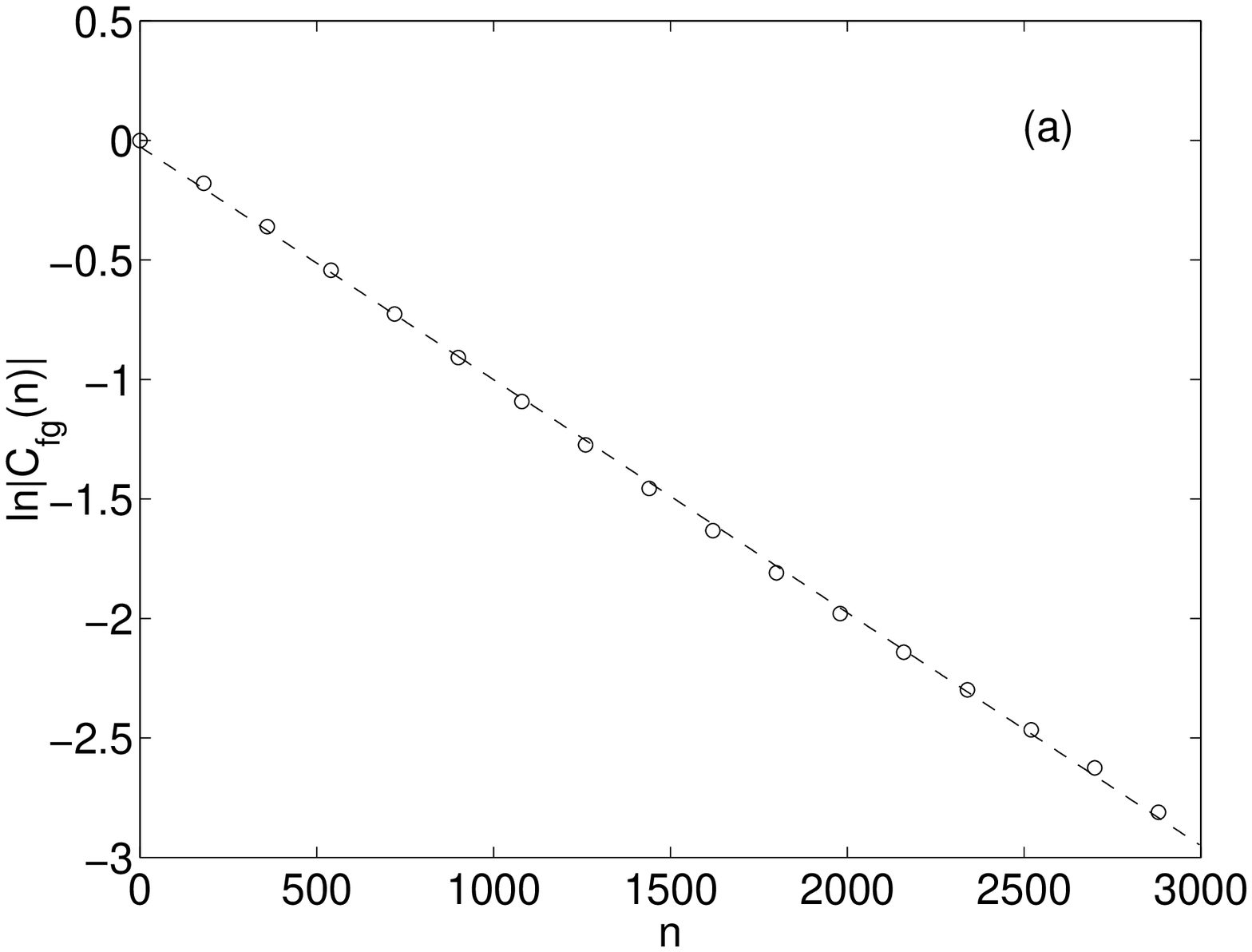} }
\end{minipage}
\hspace{1.0cm}
\begin{minipage}{7.01cm}
\centerline{\epsfxsize=7.0cm\epsfbox{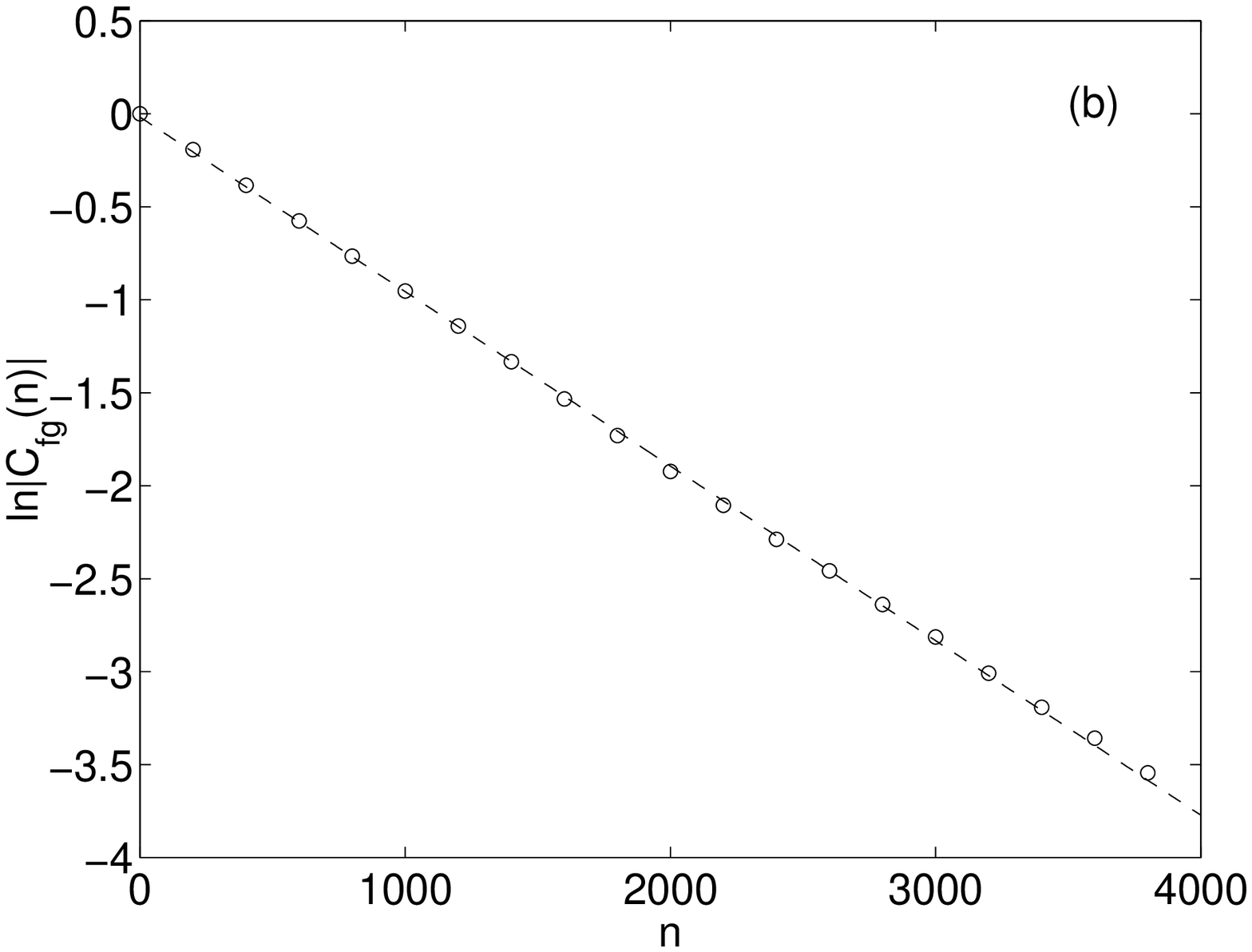} }
\end{minipage}
\\ 
\begin{minipage}{7.01cm}
\centerline{\epsfxsize=7.0cm\epsfbox{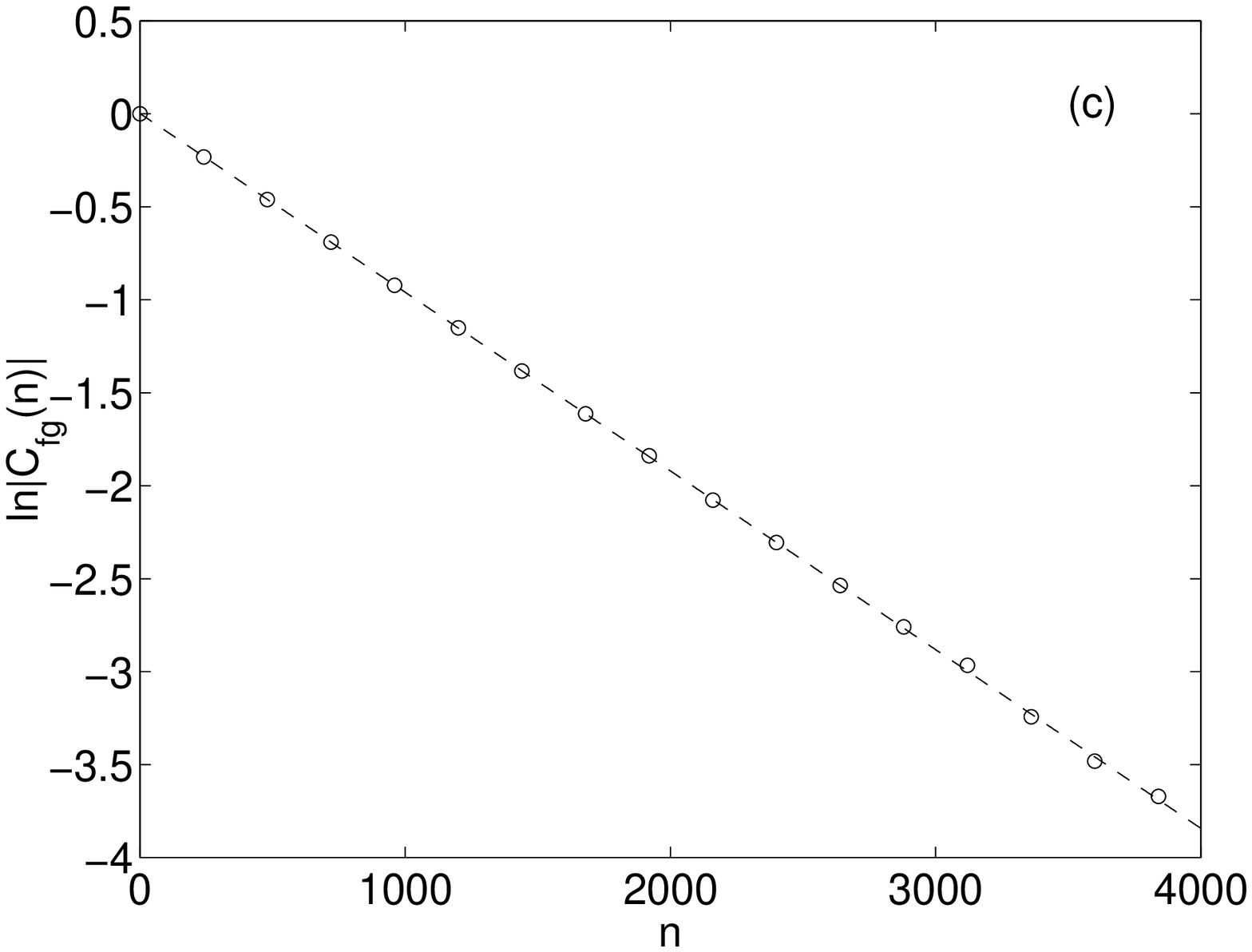} }
\end{minipage}
\hspace{1.0cm}
\begin{minipage}{7.01cm}
\centerline{\epsfxsize=7.0cm\epsfbox{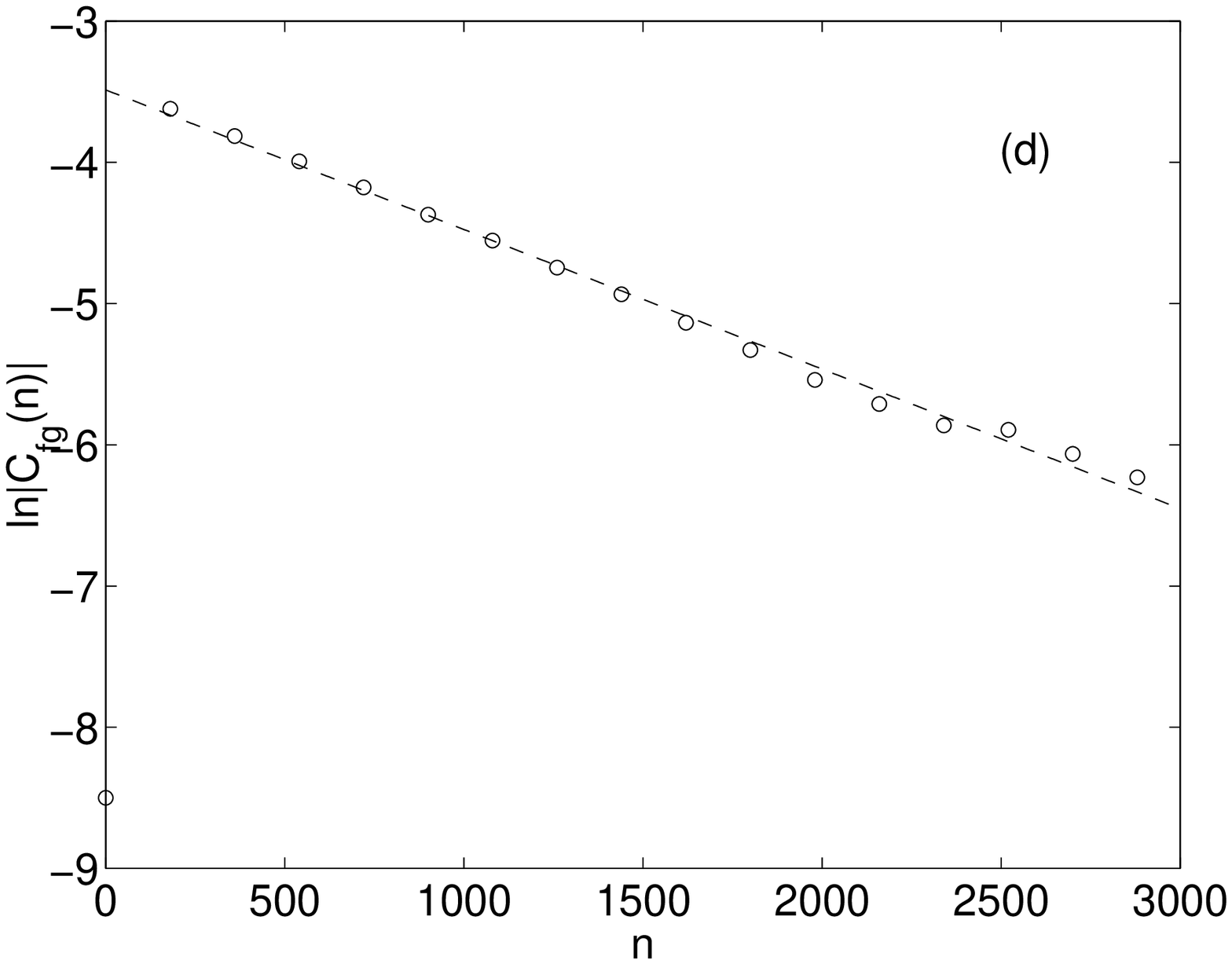} }
\end{minipage}
\end{center}
\begin{centering}
\caption{\label{max1} The function $C_{fg}(n)$ for: (a) $K=20$ ; (b) $K=30$ ; (c) $K=40$ ; (d) $K=27$, for various functions $f$ and $g$ and for various values of $s$ (Fig. 1 of \protect\cite{rot}).}
\end{centering}
\end{figure}

\begin{figure}[tb]
\begin{centering}
{\includegraphics[height=6.5cm,width=8cm]{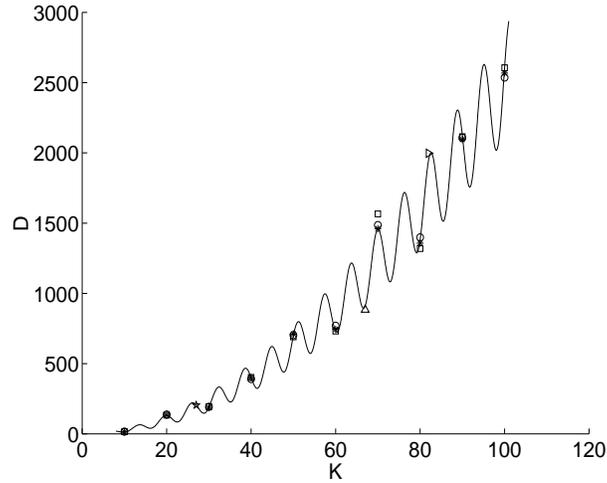}}
\caption{\label{max2} The diffusion coefficient $D$ for $K \geq 10$ as found from plots like the ones presented in Fig. 9. Various symbols represent results found for different modes $k$ while the solid line represents the analytical value (\ref{dkr4}) (Fig. 2 of \protect\cite{rot}).}
\end{centering}
\end{figure}

\begin{figure}[tb]
\begin{center}
\begin{minipage}{7.1cm}
\centerline{\epsfxsize 7.0cm \epsfbox{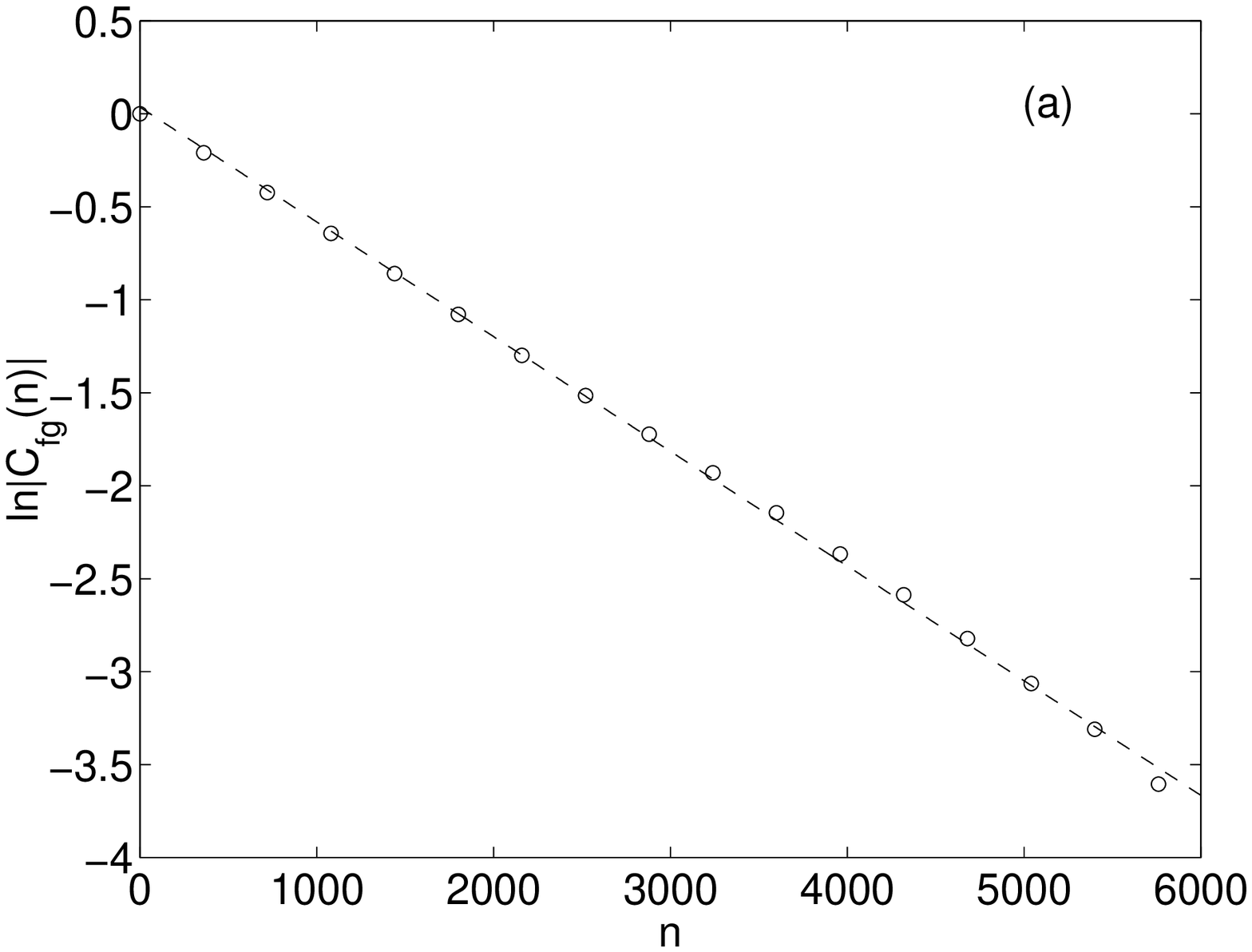} }
\end{minipage}
\hspace{1.0cm}
\begin{minipage}{7.1cm}
\centerline{\epsfxsize 7.0cm\epsfbox{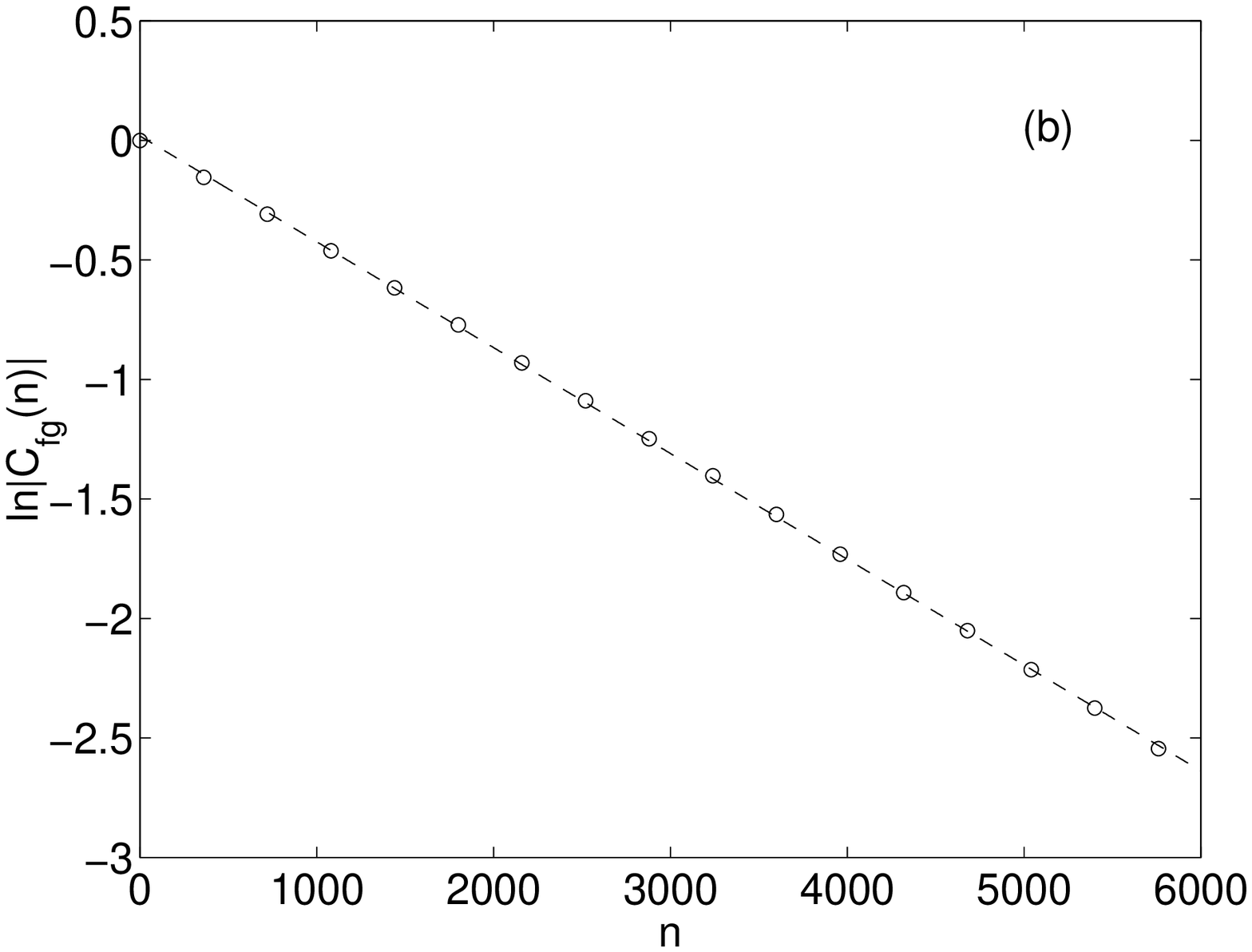} }
\end{minipage}
\\ 
\begin{minipage}{7.1cm}
\centerline{\epsfxsize 7.0cm\epsfbox{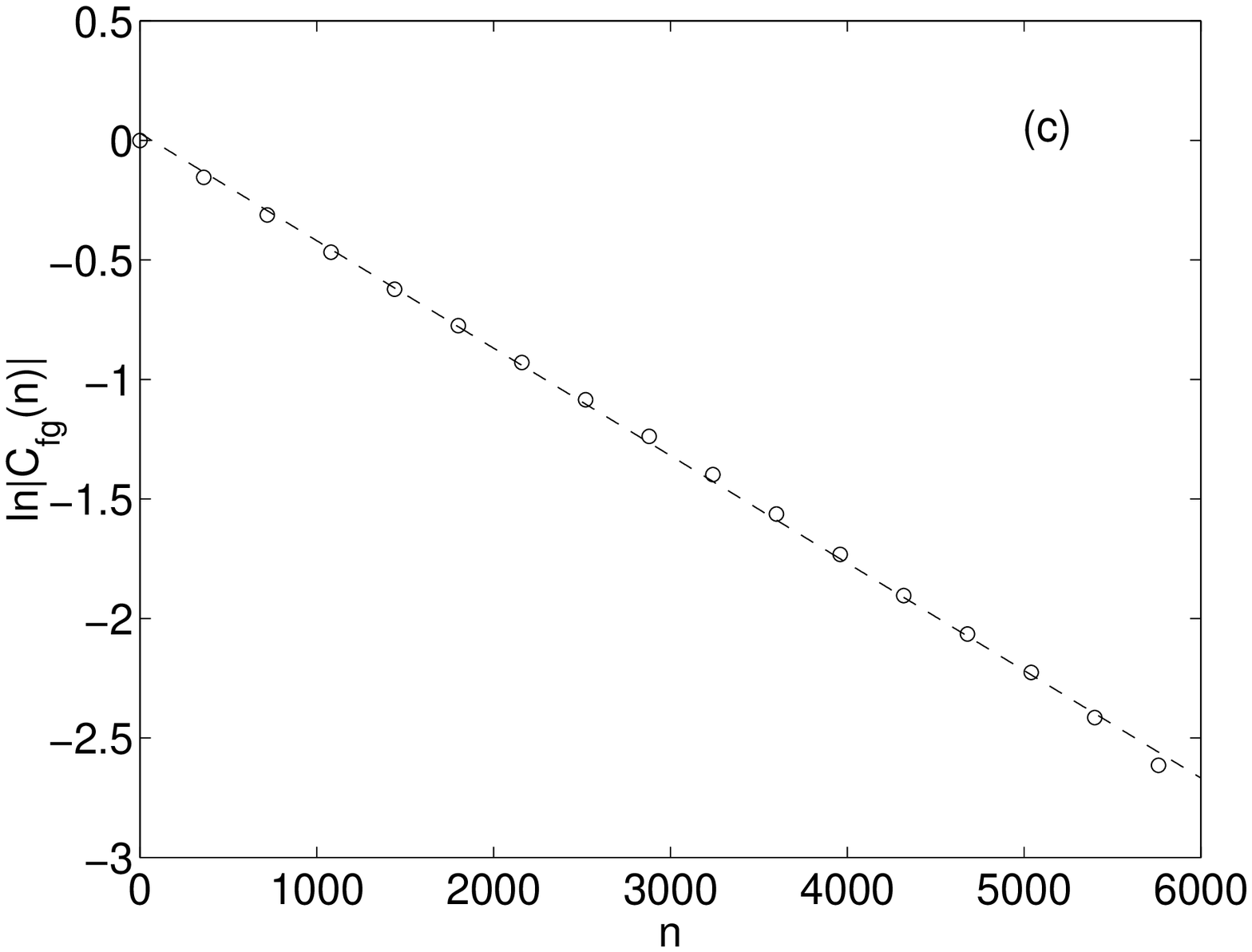} }
\end{minipage}
\hspace{1.0cm}
\begin{minipage}{7.1cm}
\centerline{\epsfxsize 7.0cm\epsfbox{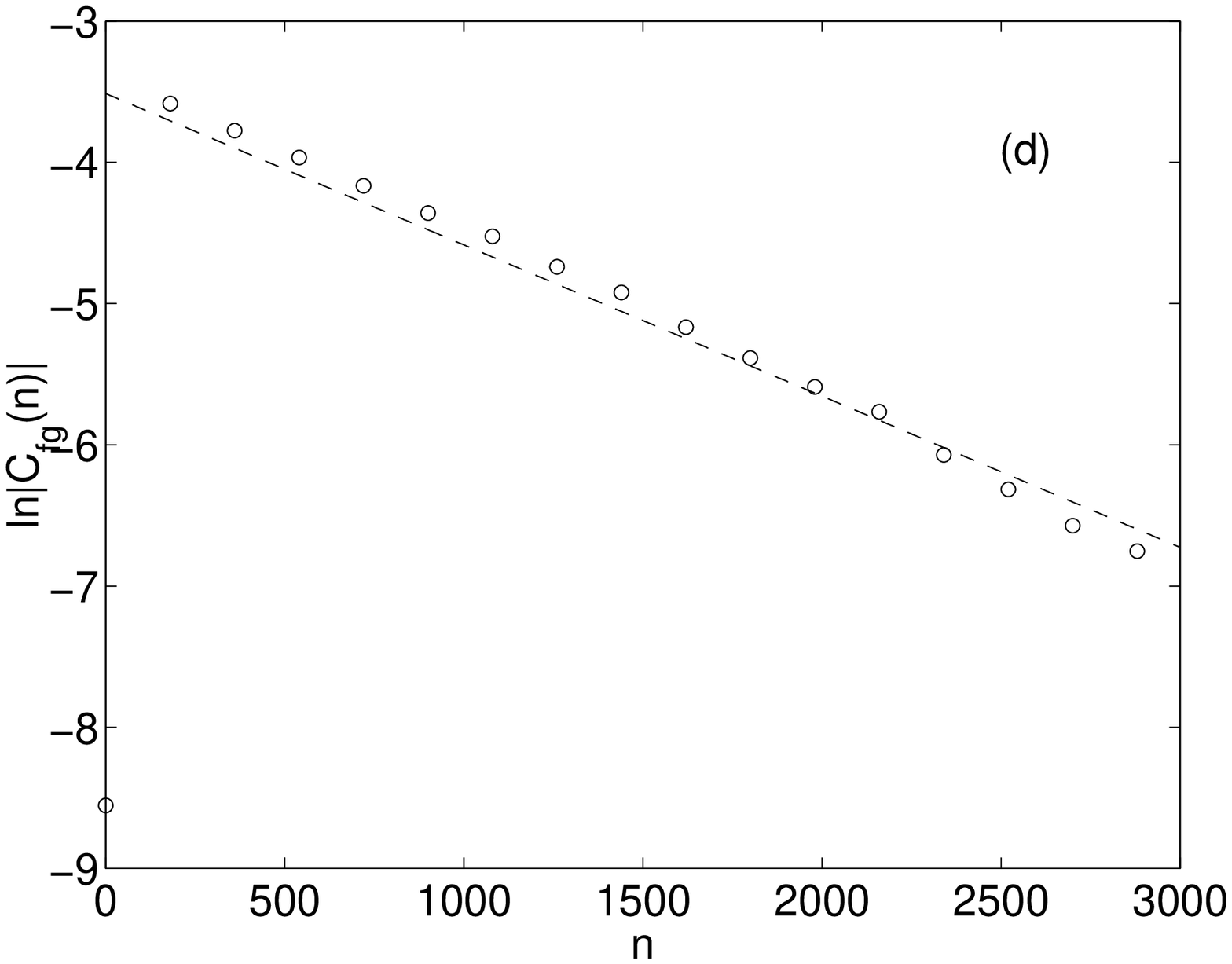} }
\end{minipage}
\hspace{1.0cm}
\end{center}
\begin{centering}
\caption{\label{max3} The function $C_{fg}(n)$ for: (a) $K=7$ ; (b) $K=8$ ; (c) $K=3$ ; (d) $K=17$, for various functions $f$ and $g$ and for various  values of $s$ (Fig. 3 of \protect\cite{rot}).}
\end{centering}
\end{figure}

\begin{figure}[tb]
\begin{centering}
{\includegraphics[height=6.5cm,width=8cm]{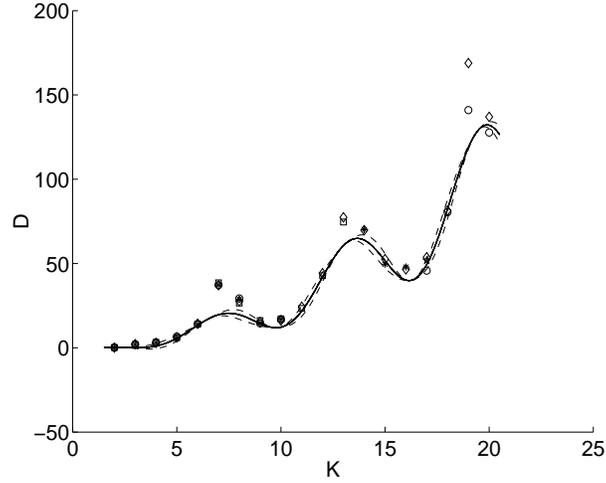}}
\caption{\label{max4} The diffusion coefficient $D$ for $K \leq 20$ as found from plots like the ones presented in Fig. 11. Various symbols represent results found for different modes $k$ while the solid line represents the analytical value (\ref{dkr4}). The dashed line represents the approximate error, resulting of the truncation of the perturbation theory expansion. The values of $D$ obtained by direct simulation of propagation of trajectories are marked by diamonds (Fig. 4 of \protect\cite{rot}).}
\end{centering}
\end{figure}

\begin{figure}[tb]
\begin{center}
\begin{minipage}{6.8cm}
\centerline{\epsfxsize 6.7cm \epsfbox{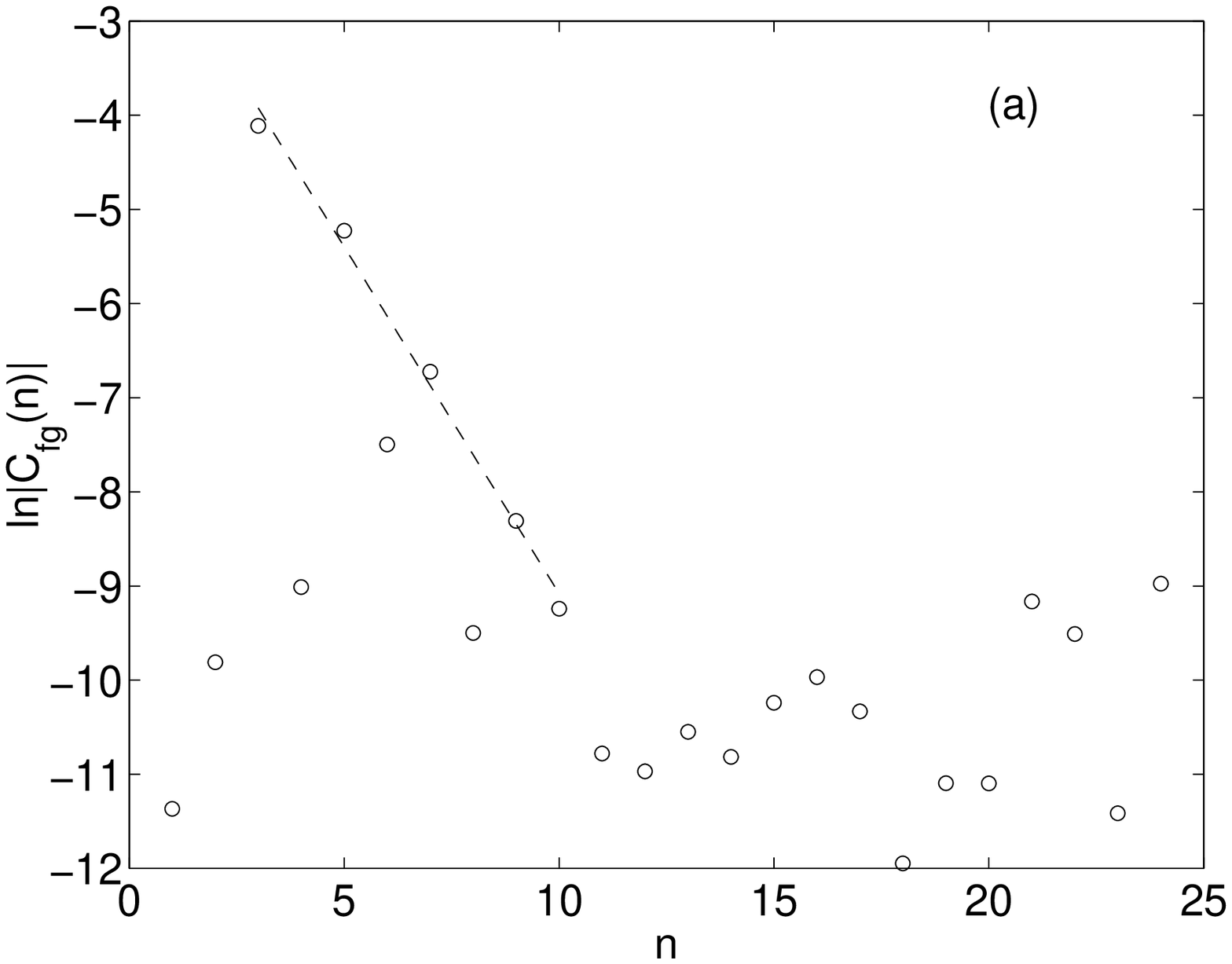} }
\end{minipage}
\hspace{1.0cm}
\begin{minipage}{6.8cm}
\centerline{\epsfxsize 6.7cm\epsfbox{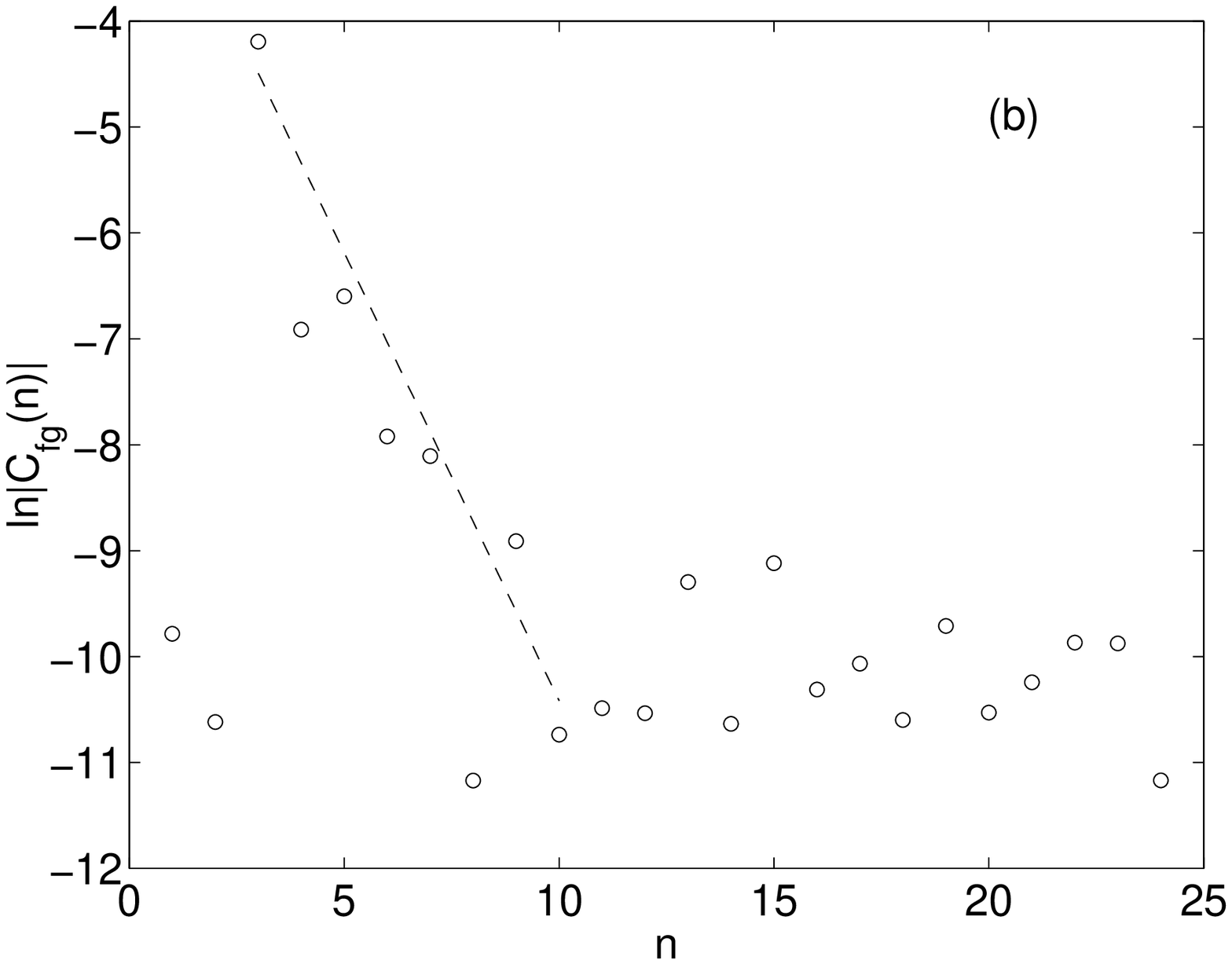} }
\end{minipage}
\\
\begin{minipage}{6.8cm}
\centerline{\epsfxsize 6.7cm\epsfbox{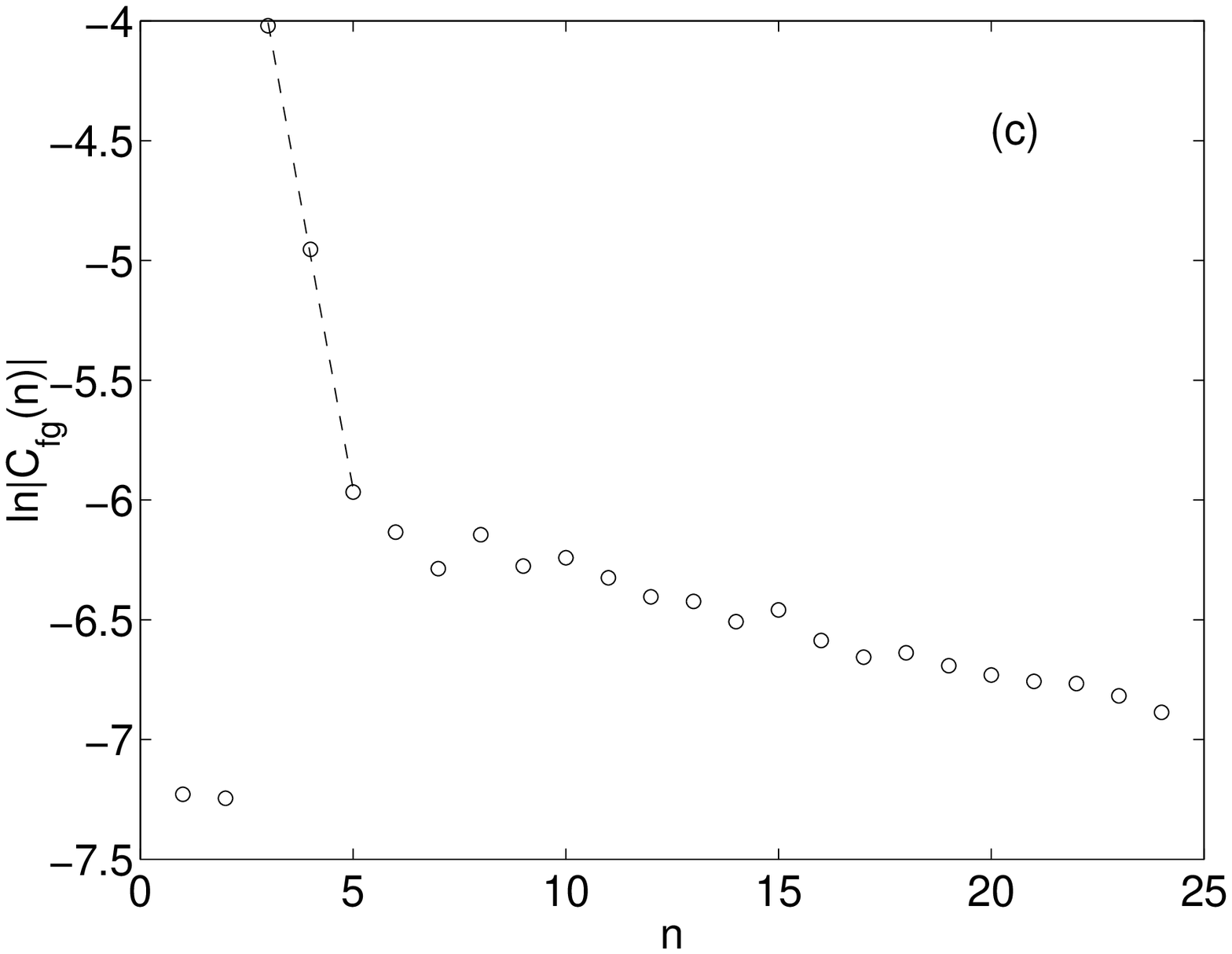} }
\end{minipage}
\hspace{1.0cm}
\begin{minipage}{6.8cm}
\centerline{\epsfxsize 6.7cm\epsfbox{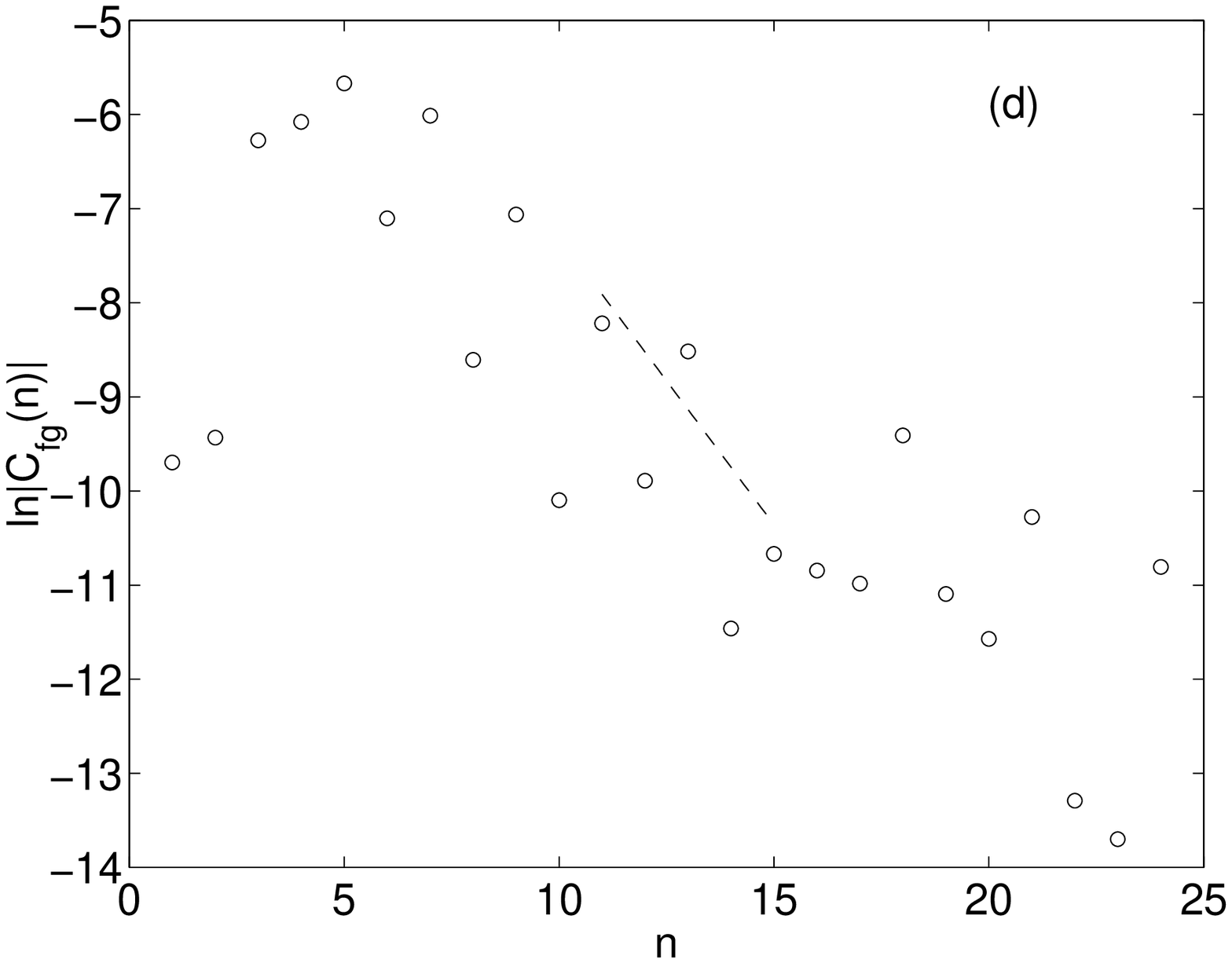} }
\end{minipage}
\hspace{1.0cm}
\end{center}
\begin{centering}
\caption{\label{max5} The function $C_{fg}(n)$ for $s=1$ with $|f \rangle=|01 \rangle, |g \rangle=|02 \rangle$ and (a) $K=16.3$; (b) $K=19.5$; (c) $K=12$; (d) $K=16$. The dashed line represents the best fit to the data (Fig. 5 of \protect\cite{rot}).}
\end{centering}
\end{figure}

\begin{figure}
\begin{centering}
{\includegraphics[height=7cm,width=8cm]{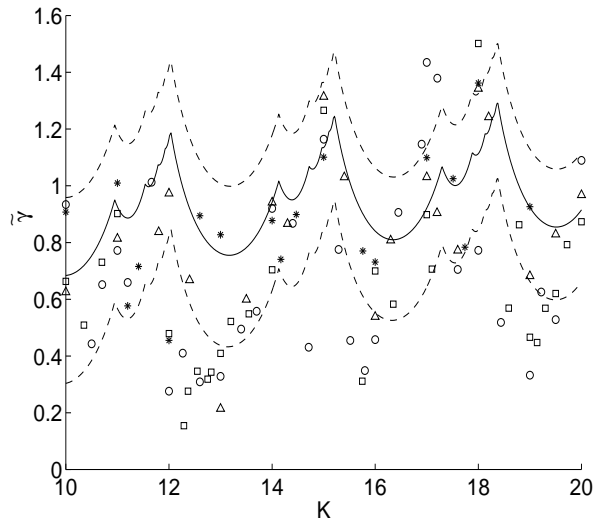}}
\caption{\label{max7} The fast relaxation rates $\tilde{\gamma}$ as found from plots like Fig. \protect\ref{max5} for various functions $f$ and $g$ with $k=0$, marked by various symbols. The analytical value (\ref{krc5}) is represented by a solid line while the dashed lines denote the analytically  estimated error, resulting of the truncation of the perturbation theory expansion (Fig. 7 of \protect\cite{rot}).}
\end{centering}
\end{figure}

In order to test the analytical results, the correlation function of the form (\ref{cor2})
\begin{equation}
\label{krn1}
C_{fg}(n)=\langle f|\hat{U}^{n}|g \rangle,
\end{equation}
where $f$ and $g$ are taken to be basis states of (\ref{fpr2}), was computed numerically. The results are plotted in Fig. \ref{max1}. 
Clear exponential decay is found. From the slopes, the relaxation rates $\gamma_k$ are obtained for the various modes,
$k$. The values of $D$ implied by $\gamma_k$ are calculated from (\ref{krc1}) and  compared to the ones obtained from (\ref{dkr4}) in
Fig. \ref{max2}. 
Good agreement was found. In Figs. \ref{max1} and \ref{max2} large values of the stochasticity parameter, $K \geq 10$, were
used. In this regime the regular regions are extremely small (they are invisible in Fig. \ref{standard}), therefore the theory presented in this
review is expected to work very well, as is indeed found in Fig. \ref{max2}.  In Fig. \ref{max3}  the correlation function
(\ref{krn1}) is presented for $K \leq 20$. 
From these results the relaxation rates $\gamma_k$ were calculated. The resulting values
of the diffusion coefficient obtained from (\ref{krc1}) are compared to the ones found from (\ref{dkr4}) in Fig. \ref{max4},
for $K \leq 20$. Here, in contrast to the regime of large $K$ presented in Fig. \ref{max2}, appreciable deviations are found. 
These are
related to sticking to regular structures, that is not taken into account in the theory presented in this review. To test the decay of correlations in the angle variable, the correlation function
(\ref{krn1}) was computed for $|f \rangle$ and $|g \rangle$ that are basis states of the form $|k=0,m \rangle$, and some of the results are presented in
Fig. \ref{max5}. 
The relaxation is very fast, therefore it is very difficult to estimate numerically the relaxation
rate.  The numerical results are compared to the analytical prediction (\ref{krc5}) in Fig. \ref{max7}. 
The analytical formula
provides a reasonable estimate of the relaxation rate and of its dependence on the stochasticity parameter $K$. There are however
deviations that are significant.  Also here some of the most significant deviations were found to be related to regular structures.

The main results that were obtained for the kicked rotor are:

\begin{enumerate}

\item Addition of noise results in the effective truncation of the Frobenius-Perron operator. In the limit of vanishing noise the 
Ruelle-Pollicott resonances are found. The resonances depend on the noise, and therefore on the truncation, as is the case for the 
kicked top, and approach limiting values inside the unit circle in the limit of vanishing noise, except the unit eigenvalue 
corresponding to the equilibrium density. 

\item The Ruelle-Pollicott resonances determine the decay of correlations. 

\item Deviations from exponential decay of correlations may be found after some time as a result of the existence of regular 
structures.

\item The following physical picture emerges. First the correlations in the angle variable $\theta$ decay. Then the correlations in 
angular momentum $J$ decay as for a true diffusion process. For finite $s$ a uniform equilibrium density is reached. For $s=\infty$, 
after the decay of the angular correlations, diffusive spreading takes place. 

\item At some time, that is increasing with the stochasticity parameter $K$, sticking to regular structures becomes important, and
the picture based on Ruelle-Pollicott resonances and exponential decay of correlations, breaks down.

\end{enumerate}

\section{Summary}

It was demonstrated that Ruelle-Pollicott resonances are relevant even when the conditions for the  Ruelle-Pollicott theorem do not 
hold. A method for the calculation of these resonances by a finite truncation of the  Frobenius-Perron operator was presented. The 
relation between the resonances and the decay of correlations was established
under these conditions. For mixed systems (as well as for other nonhyperbolic 
systems) the picture breaks down after some time, that may be very long, as a result of sticking to regular structures. 

In order to evaluate the relevance of the  Ruelle-Pollicott resonance picture for realistic systems it is instructive to introduce 
several time scales:
\begin{enumerate}

\item If $\gamma_1$ is the slowest decay rate (corresponding to the resonance that is closest to the unit circle), 
$t_{chaos}=1/{\gamma_1}$ is the decay time of correlations and the relaxation time to the invariant density.

\item The time scale when sticking to regular structures becomes important will be denoted by $t^*$.

\item In presence of external noise correlations are destroyed on the time scale $t_c \sim 1/\sigma^2$, where $\sigma^2$ is the 
variance of the noise.

\end{enumerate}

In order to observe the exponential decay of correlations it is required that
\begin{equation}
\label{s1}
t_{chaos} \ll~t_c,~t^*.
\end{equation}
Destruction of sticking to regular structures requires $t_c<t^*$. In the regime 
\begin{equation}
\label{s2}
t_{chaos} \ll~t_c<~t^*
\end{equation}
the decay of correlations is expected to be similar to the one of hyperbolic systems and one should be able to explore it in the 
framework of the method that was outlined in the present review. Application of the method to specific examples is of great 
interest.

\acknowledgments
It is our great pleasure to thank Fritz Haake for many illuminating discussions during the preparation of the lectures and of the review, 
for the files of the figures from \cite{haake} and for the permission to use them.  Part of the review is based on work in collaboration with Oded Agam and
Maxim Khodas, that is acknowledged with great pleasure. We would like to thank Christopher Manderfeld and Joachim Weber for useful
discussions and communications.  SF thanks Andreas Buchleitner for the hospitality at the Max Planck Institute for the Physics of Complex
Systems in Dresden, where the lectures were prepared.  This research was supported in part by the US-Israel Binational Science Foundation
(BSF), by the Minerva Center of Nonlinear Physics of Complex Systems, by the Max Planck Institute for the Physics of Complex Systems in Dresden,
and by the fund for Promotion of Research at the Technion.


Most of the results presented in the review can be found in references marked by **
\end{document}